\pdfoutput=1  
\documentclass[a4paper,fleqn]{cas-sc}

\usepackage[numbers,sort&compress]{natbib}
\usepackage{amsmath,amssymb}
\usepackage{booktabs}
\usepackage{array}
\usepackage{rotating}
\usepackage{enumitem}
\usepackage{textcomp}
\usepackage{float}
\usepackage[section]{placeins}

\usepackage{xcolor}
\usepackage{listings}
\definecolor{codebg}{rgb}{0.97,0.97,0.97}
\graphicspath{{figures/}}

\ExplSyntaxOn
\cs_set:Npn \__first_footerline: {}
\ExplSyntaxOff

\begin{document}
\let\WriteBookmarks\relax

\shorttitle{Systematic Multi-Agent AI from Advanced Regulatory Control}
\shortauthors{I. B. R. Nogueira and S. Skogestad}

\title[mode=title]{A Systematic Approach to Multi-Agent AI from Advanced Regulatory Control Theory: Safe and Auditable LLM Operator Agents for Process Control}

\author[1]{Idelfonso B. R. Nogueira}
\cormark[1]
\ead{idelfonso.b.d.r.nogueira@ntnu.no}

\affiliation[1]{organization={Department of Chemical Engineering, Norwegian University of Science and Technology (NTNU)},
            city={Trondheim},
            postcode={7491},
            country={Norway}}

\author[1]{Sigurd Skogestad}

\cortext[1]{Corresponding author}

\begin{abstract}
Recent literature shows that large language models (LLMs) are useful for
general-purpose tasks yet perform poorly on specific domain ones. One reason
is the difficulty of supplying narrow context to a general-purpose model and
of bounding the task it is asked to perform. It is possible to hypothesise
that a multi-agent reformulation under process-control principles offers a
route to address those points, since control theory provides a discipline of
decomposing a system into elements of contained scope, each defending one
controlled variable, with conflicts resolved by structural priority: MIN/MAX
selector networks for CV--CV switching and split-range (split-parallel) logic
for MV--MV switching. The present work proposes such a reformulation, derived
from Advanced Regulatory Control (ARC) theory. Each
feedback loop in the ARC chain is mapped to one specialised LLM operator
agent carrying the loop's control-theoretic context (controlled variable,
setpoint, chain priority, selector kind). The chain's interaction logic
(MIN/MAX selectors, override paths) is encapsulated as a single orchestrator
agent. Two orchestrator variants are tested: a deterministic rule chain,
and a Claude-based LLM orchestrator at a slower tier. The control principles
limit each agent's task and inform how its limitations are handled. The
multi-agent system inherits the safety property of the ARC chain: every
constraint conflict is resolved deterministically by the orchestrator,
regardless of the LLM output. Evaluated on a dairy-barn ventilation case
over a 4-day mixed-season scenario, Qwen 2.5 7B Instruct operator agents
running offline on a 24\,GB consumer GPU at a 5-minute cadence produce
auditable trajectories, each paired with an operator-voice rationale that
supports a control campaign logbook.
\end{abstract}


\maketitle

\section{Introduction}
\label{sec:intro}

Process systems engineering has long approached the control of complex
multivariable plants by decomposing the problem into a hierarchy of simpler
interacting pieces~\cite{skogestad2000plantwide,baldea2012,luyben1998}. At the
regulatory layer, this decomposition is realised through cascade structures
(a master loop setting the setpoint of a faster slave loop) and selector and
split-range structures that allow each feedback loop to defend a single controlled
variable, with priority among loops resolved by MIN/MAX
selector networks when several controlled variables compete for one
manipulated variable (CV--CV switching) and by split-range logic when one
manipulated variable must be shared in a priority order among loops (MV--MV
switching)~\cite{maarleveld1970,shinskey1981}. At the supervisory layer,
the same discipline produces the well-known plant-wide control hierarchy, in
which real-time optimisation, supervisory regulation, and basic control occupy
distinct time-scales and communicate through a small number of setpoint
handoffs~\cite{skogestad2000plantwide,engell2007}. More recent work extends
this framework to operating regions in which the active set of binding
constraints changes with time, producing the Advanced Regulatory Control (ARC)
pattern: a network of feedback loops, MIN/MAX selectors for CV--CV switching,
and split-range arrangements for MV--MV switching, which together resolve
constraint conflicts by structural priority rather than by online
re-optimisation~\cite{skogestad2023arc,skogestad2026arc,reyes2020,reyes2019,krishnamoorthy2022,forsman2025splitparallel}.
A related primitive, mid-ranging, is distinct in purpose: it uses a secondary
manipulated variable to assist a primary one for dynamic (speed-of-response)
reasons, not to resolve a constraint conflict.
In the language of modern artificial intelligence, it is possible to read
every selector-based control structure as a multi-agent system: each loop can
be seen as an agent that owns one controlled variable, publishes a candidate
manipulated-variable value, and yields to higher-priority agents through the
selector network.
This reading is presented here as a hypothesis, since the equivalence has not
been formally established in the literature. The vocabulary used for several
decades by the process-control community (coupling, pairing, anti-windup,
bumpless transfer) can be read, under this hypothesis, as a vocabulary of
agent decomposition.

The reverse reading is equally productive: multi-agent systems built from large
language models (LLMs) have, over the past three years, developed a design
vocabulary that translates directly into the process-control idiom. In financial
decision-making, frameworks such as
FinCon~\cite{fincon2024} and TradingAgents~\cite{tradingagents2025} orchestrate
teams of specialist agents that share verbal reinforcement and collectively
outperform single-model baselines on portfolio and trading tasks. In business
administration, hierarchical loan-decision and compliance pipelines deploy
supervisor and worker agents to handle multi-step workflows that are
impractical for a single model; the patterns that emerge across these settings
have been catalogued in recent taxonomies of hierarchical multi-agent
design~\cite{hierarchical_mas_taxonomy2025}. In software engineering and research,
agent-team patterns built on chain-of-thought
reasoning~\cite{wei2022chain,yao2022react,huang2022language} and on
generative-agent simulations~\cite{park2023generative,wooldridge2009,russell2020}
have produced a stable design vocabulary of orchestration, append-only
workspaces, schema-validated inter-agent messages, and structured outputs.
It is possible to import this vocabulary into process control directly, and
the present work does so, borrowing the orchestrator+advisor pattern from
adjacent fields and grounding it in the same deterministic chain that
today's ARC networks already use.

Despite the maturity of multi-agent systems in adjacent fields, the systematic
application of multi-agent architectures to process systems engineering remains
sparse. The regulatory layer has been largely untouched.
Early agent-based work in the chemical industry concentrated on
plant-wide process supervision and ontology-driven coordination across
distributed units~\cite{natarajan2013mas} rather than on the regulatory control
layer itself. The recent wave of LLM-agent work in chemical
engineering, summarised in the review of multi-agent systems for chemical
engineering~\cite{mas_che_review2025} and in the contemporary work on
specialised multi-agent LLMs for process systems
engineering~\cite{pse_specialized_mas2026,costa2026maasc}, focuses largely on modelling,
simulation, and design tasks rather than on closed-loop regulatory control.
Contemporary work on autonomous industrial control with LLM
agents~\cite{industrial_llm_agentic2024} has explored configurations in which
the language model writes control actions at the inner loop, sometimes
generating manipulated-variable commands directly from the process state and
sometimes supervising a numerical controller. Those configurations show that
LLM agents are capable of producing operator-style decisions in industrial
settings, and they inform two design questions: where in the control hierarchy
the language model should sit, and how its outputs should be bounded so that
safety properties of the deterministic layer are preserved. The position
adopted in the present work places the LLM operator agents at the regulatory
tier behind a structural chain, and the orchestrator at the supervisory
tier. From this perspective, a gap is identified at a specific point: how
should a multi-agent regulatory layer be constructed using the toolkit that
process systems engineering has already developed for the same purpose?

The specialisation of agents in multi-agent large language model systems is
commonly achieved through retrieval-augmented generation (RAG). In RAG, the
agent's context is enriched at inference time by retrieving relevant passages
from a document store and embedding those passages in the prompt alongside
the query. Fine-tuning represents an alternative route to specialisation; its
computing-power requirements are substantial, however, and replicating a
fine-tuned model for every specific industrial task is not feasible with
current technology. RAG therefore remains the practical option for most
deployments. The evaluation of embedding-based retrieval as a reliable
specialisation mechanism is rarely done against worst-case guarantees.
Furthermore, it was not found any theoretical result ensuring that the
context supplied by a RAG system will contain the information the model
actually needs: the probability of a critical passage being omitted is not
negligible, particularly when that information is idiosyncratic or appears
with low frequency in the corpus~\cite{weller2025rag}. For process systems
engineering applications, those are important points to be addressed. The
decisive information that distinguishes one process unit from another resides,
typically, in its peculiarities: the non-standard operating conditions, the
site-specific constraints, the failure modes documented in field notes rather
than textbooks. Those peculiarities are precisely the passages most likely to
be displaced during retrieval by semantically dominant but operationally less
relevant material. A multi-agent architecture is capable of addressing this
limitation through a different route. By decomposing the control problem into
elements of concise and contained scope, it is possible to assign each agent
a single, well-defined task whose context is narrow enough to be fully
specified in a short, deterministic prompt. The context required by any one
agent is limited by construction, and the retrieval problem that RAG must
solve is replaced by a prompt engineering problem that a domain expert can
verify by inspection. Advanced regulatory control theory provides exactly
this kind of decomposition. The plant-wide control procedure produces a
roster of specialised feedback loops, each defending one controlled variable
with one manipulated variable, each requiring only its own measurement,
setpoint, and selector priority in order to act. The controlled variable a
loop defends is often an inequality constraint, but it need not be: a loop may
equally hold a floating setpoint that is itself the output of a higher loop. As it will be demonstrated in this work,
the correspondence between the ARC decomposition and the agent context
requirement is the foundation of the multi-agent architecture proposed here.

The question motivates a two-way bridge that the present work attempts to
construct. In the first direction, from process control theory to multi-agent
systems, it is possible to recognise that the architectural primitives required
to build a well-behaved multi-agent regulatory layer already exist inside
classical control: the variable-coupling logic of the selector chain defines the
message-passing topology; the MIN/MAX selector network is the priority
orchestrator; the back-calculation anti-windup of \AA{}str{\"o}m and
H{\"a}gglund~\cite{astrom-pid} is the mechanism behind what we term bumpless
agent re-engagement, borrowing the standard ARC term \emph{bumpless transfer}
for an agent that resumes authority without a step in its manipulated variable;
and a safety override is not a separate non-selector guard but the final,
most-critical element of the chain itself, the selector that the others yield
to when constraints conflict. In the
second direction, from multi-agent artificial intelligence to process control,
the contribution is that the language model can also act as a slow gain scheduler
that retunes the orchestrator's reasoning parameters in response to the current
operating regime, while leaving the deterministic numerical control chain in charge
of every manipulated variable. This proposed architecture can place the language
model at the operator tier: the model receives the current state of its assigned
controlled variable and produces a mode classification and a rationale, while the
lower control chain retains authority over the manipulated variables. This sort of
implementation places the language model also at the supervisory tier of the
classical hierarchy. If one uses a time-scale separation of approximately two
orders of magnitude relative to the regulatory loop, a separation of the order
conventionally taken to mark the boundary between the supervisory and regulatory
layers of the classical hierarchy~\cite{skogestad2000plantwide}, the system is
protected from possible adverse advice that, in this way, can only nudge the
closed-loop behaviour, not destabilise it within one advisor period.

Three specific contributions support the bridge. The first is the
demonstration that a multi-agent reformulation of the ARC selector chain
produces trajectories indistinguishable from those of the ARC network when the
operator agents' decision logic is implemented as a SIMC-tuned PI block,
formalising the correspondence between classical selector control and
structured agent broadcast. This validation swap isolates any subsequent
behavioural difference as attributable to the operator agents themselves or
to the orchestrator's added capabilities, not to the multi-agent
decomposition itself. The second contribution is the construction of three
architectures from the same ARC chain. Architecture~A is the PI
baseline, which is the ARC network itself. Architecture~B and
Architecture~C share a common LLM-operator implementation, in which each
of the six operator agents is backed by a local Qwen 2.5 7B Instruct call,
and they differ only at the orchestrator tier: Architecture~B uses a
deterministic rule-based orchestrator that applies the MIN/MAX priority
chain without LLM involvement, while Architecture~C adds a Claude
Opus 4.7 LLM-based orchestrator that classifies the operating regime and
tunes the orchestrator's dwell-time, trend-bias, and memory parameters on
a slow supervisory tier. In both architectures the structural priority
chain is preserved deterministically and the LLM cannot override the
chain's binding-direction decisions. The agent roster is derived from
the control-theory-designed chain through two equivalences: each PI in
the chain corresponds to one operator agent that carries
control-theory context (CV, MV, setpoint, priority, SIMC tuning,
anti-windup) and acts directly on its proposal target; the chain's
interaction logic (MIN/MAX selectors, split-parallel, overrides) is
encapsulated as one orchestrator agent. It is possible to see the
multi-agent reformulation as a transcription of the plant-wide control
procedure into agents, with the orchestrator-tier LLM operating as a
parameter scheduler in the spirit of the classical supervisory layer. The third contribution is that conflict resolution
among agents follows structural priority, not negotiation or online
optimisation. The orchestrator agent encodes the priority order prescribed
by the ARC chain: every constraint conflict is resolved deterministically,
without LLM involvement. It is possible to see this property as the
agent-level realisation of Skogestad's MIN/MAX selector chain
\cite{skogestad2026arc,maarleveld1970,shinskey1981}; it is also what supports the deployment of LLM
agents in real process control systems with structural soundness: an LLM
output that misclassifies the regime can only shift the proposal magnitude,
since the chain's priority order is enforced by the orchestrator regardless.
A second consequence of the agent-based reformulation is that the SIMC PI
tuning step of the classical recipe is replaced by a different design step.
The LLM operator agent does not require numerical gain tuning, since it
does not produce a numerical action through $K_c \cdot e$; it produces a
regime classification that the agent code maps to a proposal. What the
operator agent does require is a curated prompt and a curated context, so
that the model is informed of its role in the chain, the selector kind it
serves, the binding direction, and the rolling history of its own past
decisions. The PID tuning step of classical ARC is, in this sense, replaced
by a prompt engineering and context curation step, which is described in
detail later in the paper.

The two contributions above are evaluated empirically on the cow-barn
ventilation case introduced in~\cite{skogestad2026arc}, a process in which freeze
protection and CO\textsubscript{2} comfort become physically incompatible
during deep winter. The case study is the evaluation vehicle, not a
contribution of its own. The ARC chain is taken as the reference
architecture and a multi-agent system is constructed under the recipe above.
The headline architecture (Architecture~C) assigns to each PI's role in the
chain one operator agent backed by a local instruction-tuned LLM
(Qwen~2.5 7B Instruct, running offline on a single consumer GPU), and a
Claude-based LLM orchestrator at the supervisory tier that tunes the
chain's parameters while the structural MIN/MAX priority order remains
enforced deterministically. Over a 4-day mixed-season scenario (winter cold
combined with summer warm in 96 hours) the LLM-MAS is capable of producing
structurally correct operator-style decisions at a 5-minute polling cadence,
with a narratable audit trail (one operator-voice rationale per agent per
sample): two properties that the monolithic ARC implementation cannot offer.

The remainder of this paper is organised as follows.
Section~\ref{sec:case-cow} presents the cow-barn case study: the plant
model, the 4-day mixed-season disturbance scenario, the construction of
the multi-agent architecture from the ARC chain, the design of the LLM
operator agents and their context, the orchestrator agents, and the
empirical evaluation of the three architectures.
Section~\ref{sec:impl} collects the implementation notes common to all
architectures and the agent-system limitations exposed by the multi-agent
reformulation. Section~\ref{sec:conclusions} closes with the conclusions of the present evaluation.

\section{Case study: happy-cow barn}
\label{sec:case-cow}

The architecture proposed in this work follows from the plant-wide
control procedure of Skogestad~\cite{skogestad2000plantwide,
skogestad2026arc}: apply advanced regulatory control principles to a
specific plant, derive the chain, and translate the chain into a
multi-agent system using the equivalences established in
Section~\ref{sec:systematize}. The cow-barn ventilation problem
introduced in Skogestad~\cite{skogestad2026arc} is
used here as one concrete instantiation of this general procedure: it
offers two manipulated variables, two competing constraints that
become physically incompatible during deep winter, and a natural
priority conflict that the MIN/MAX selector network resolves by
structural priority. Those properties make it a representative test
case for the architecture. We work the case end-to-end: process
description, plant model, plant imperfections, disturbance scenarios,
the construction of the multi-agent architecture from the ARC
chain, the design of the LLM operator agents and their context, the
orchestrator agents (rule-based and LLM-based variants), and the
empirical evaluation against the PI chain baseline.
\subsection{The happy-cow case study}

We adopt the \emph{happy-cow} livestock-housing problem introduced in
Skogestad~\cite{skogestad2026arc}. A modern dairy barn is ventilated by a fan and
warmed by an electrical heater. The CO\textsubscript{2} concentration $C$ must remain
below 1\,000~ppm whenever feasible, a \textit{comfort} upper bound, and below
3\,000~ppm at all times as a \textit{health} limit; while the temperature $T$ must lie
between 5\,\textdegree C and 20\,\textdegree C whenever feasible and above 0\,\textdegree C
at all times as a freeze-protection lower bound.
The plant offers two manipulated variables: the variable-frequency-drive
ventilation fan $u_1\in[0,100]\,\%$ (which reduces both $C$ and $T$ by replacing
inside air with outside air) and an electric heater $u_2\in[0,100]\,\%$ (which raises
$T$ only). Disturbances enter through four channels: the outdoor temperature
$T_{\text{out}}(t)$, the cow occupancy $n_{\text{cows}}(t)$, an unmeasured
CO\textsubscript{2} source $q^{\text{extra}}_{\text{CO}_2}(t)$ (bedding fermentation,
manure-pit gas), and an unmeasured envelope-leak coefficient $UA^{\text{extra}}(t)$
(a door propped open). The first two are visible to the operator and the controller;
the last two are not.

The case is interesting because it is \emph{constrained-mostly}: the bound on $T$ is two-sided,
the bound on $C$ is one-sided, and during deep winter the freeze-protection limit and the
CO\textsubscript{2}-comfort limit are physically incompatible. With one fan and one heater
in the $-15$\,\textdegree C-outside regime, no combination of $(u_1, u_2)$ keeps both
$C \le 1\,000$ and $T \ge 5$; the controller must give up one constraint to honour the other.
The ARC chain of Skogestad~\cite{skogestad2026arc} advances a particular solution: a hierarchical PID-with-selectors
network that resolves the conflict implicitly by priority order. We re-implement this
baseline and contrast it with three further architectures.
The lowest-priority input to the chain is an operator-defined desired fan
value $u_0$, set to $50\,\%$ as a representative nominal point; strictly, $u_0$ is
the unconstrained optimum of a steady-state objective trading fan electrical cost
against air quality and acoustic comfort, a problem that is not solved in this work,
though refining $u_0$ via online unconstrained optimisation, possibly driven by an
additional supervisory agent, represents a logical extension of the present architecture.

\subsection{Plant model}

\begin{figure}[ht!]
\centering
\includegraphics[width=\linewidth]{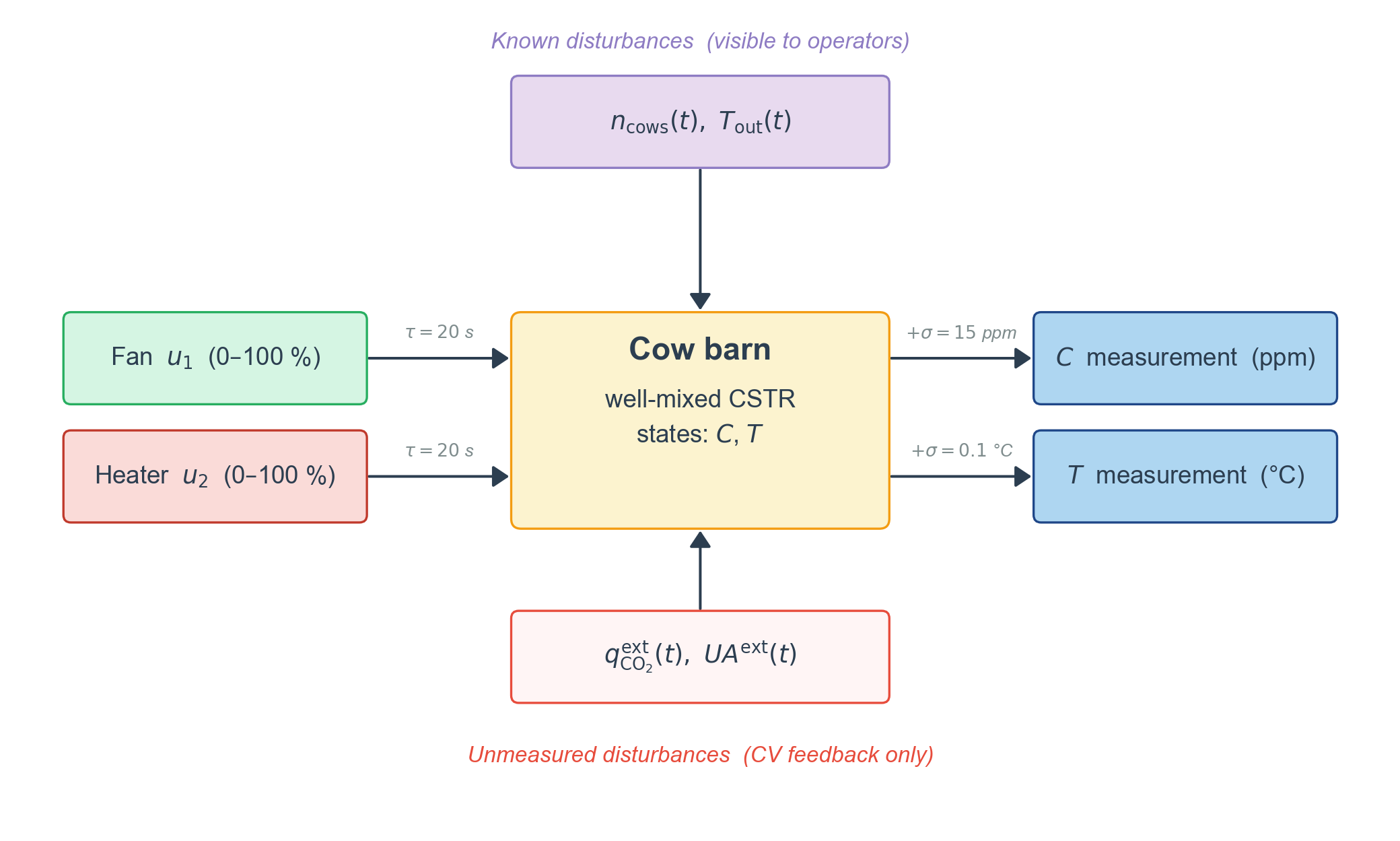}
\caption{Cow-barn well-mixed CSTR: two MVs ($u_1$, $u_2$), two CVs ($C$, $T$), four disturbance channels (two known, two unmeasured).}
\label{fig:plant}
\end{figure}

We model the barn as a single well-mixed continuous-stirred-tank reactor (CSTR)
with two states $(C, T)$, two manipulated variables, and four disturbance
channels (Figure~\ref{fig:plant}). The mass balance for CO\textsubscript{2} and the energy
balance read
\begin{align}
V\,\dot{C} \;=\;& n_{\text{cows}}\,q_{\text{CO}_2} + q^{\text{extra}}_{\text{CO}_2}(t) + Q_{\text{air}}(C_{\text{out}} - C),
\label{eq:co2balance}\\[2pt]
\rho c_p V\,\dot{T} \;=\;&
n_{\text{cows}} Q_{\text{cow}} + P_{\text{h}}\frac{u_2}{100} - \rho c_p Q_{\text{air}}(T - T_{\text{out}}) \nonumber\\
& - \bigl(UA + UA^{\text{extra}}(t)\bigr)(T - T_{\text{out}}),
\label{eq:Tbalance}
\end{align}
with the fan map
\begin{equation}
Q_{\text{air}}(u_1) \;=\; Q_{\min} + (Q_{\max} - Q_{\min})\,\frac{u_1}{100}.
\label{eq:fanmap}
\end{equation}
Numerical values used throughout this work are listed in
Table~\ref{tab:plant-params}; they were chosen as plausible defaults for a medium
Norwegian dairy barn, since the source ARC study~\cite{skogestad2026arc} states
its case qualitatively and does not report a numerical dataset. To anchor the
comparison to the source conditions rather than to our own parameter choices,
Section~\ref{sec:res-sigurd} additionally reproduces the exact plant parameters
of that study's Case~IIB (the $50$\,kW heater, $UA = 2000$\,W/K, and the
$0 \to -40 \to +15$\,\textdegree C ramp of its Table~2). Those parameters are
the ones under which the $3000$\,ppm sick-CO\textsubscript{2} limit becomes an
active constraint: the per-cow heat release and CO\textsubscript{2} production
are tied through the same combustion stoichiometry and cannot be varied
independently, so it is the deep-cold heater saturation, not a change in
occupancy, that forces the CO\textsubscript{2} ceiling to bind.

\begin{table}[!ht]
\centering
\caption{Plant parameters used in this study.}
\label{tab:plant-params}
\begin{tabular}{l l l}
\toprule
\textbf{Symbol} & \textbf{Value} & \textbf{Meaning} \\
\midrule
$V$                          & 3\,000\,m\textsuperscript{3}              & Air volume of the barn \\
$n_{\text{cows}}$ (nominal)  & 80                                       & Cow occupancy \\
$q_{\text{CO}_2}$            & $5\times10^{-5}$\,m\textsuperscript{3}/s/cow & Per-cow CO\textsubscript{2} generation ($\approx$180\,L/h) \\
$Q_{\text{cow}}$             & 1\,000\,W/cow                            & Per-cow sensible metabolic heat \\
$Q_{\max}, Q_{\min}$         & 15, 0.5\,m\textsuperscript{3}/s          & Fan airflow at 100\,\% and 0\,\% \\
$P_{\text{h}}$               & 100\,kW                                  & Heater rating at 100\,\% \\
$UA$                         & 1\,500\,W/K                              & Building heat-loss coefficient \\
$C_{\text{out}}$             & 420\,ppm                                 & Outdoor CO\textsubscript{2} concentration \\
$T_{\text{out}}$ (nominal)   & 5\,\textdegree C                         & Outdoor temperature \\
$\rho$, $c_p$                & 1.2\,kg/m\textsuperscript{3}, 1\,005\,J/kg/K & Air density, specific heat \\
\bottomrule
\end{tabular}
\end{table}

The two natural time constants at the nominal operating point ($u_1=50\,\%$) are
\begin{equation}
\tau_C = \frac{V}{Q_{\text{air}}} \approx 387\text{ s}, \qquad
\tau_T = \frac{\rho c_p V}{\rho c_p Q_{\text{air}} + UA} \approx 334\text{ s},
\label{eq:taus}
\end{equation}
both lying in the 5--7~minute range, dictating the choice of controller sample
time and tuning band.

\FloatBarrier

\subsection{Plant imperfections imposed in this study}
\label{sec:imperfections}

To stress the architectures under realistic operating conditions, the ideal plant
of (\ref{eq:co2balance}--\ref{eq:Tbalance}) is wrapped in an imperfect-plant
simulator that adds three classes of imperfection identically across all controllers:
\begin{enumerate}[leftmargin=22pt,itemsep=2pt]
  \item \emph{Measurement noise.} Independent zero-mean Gaussian noise with
        $\sigma_C = 15$\,ppm and $\sigma_T = 0.1$\,\textdegree C is added to each CV
        sample the controller receives.
  \item \emph{Actuator deadtime.} Both $u_1$ and $u_2$ are filtered through identical
        first-order lags with $\tau_{\text{act}} = 20$\,s before reaching the plant,
        representing realistic VFD ramp-up and heater contactor dynamics.
  \item \emph{Random-number seed.} All controllers see the exact same noise realisation
        (seed = 11), so any difference in their trajectories is attributable to the
        controller alone.
\end{enumerate}

\subsection{Disturbance scenario}
\label{sec:scenario}

Throughout this paper, the controllers are evaluated on a 4-day mixed-season
scenario (96 hours) designed to exercise the cold regime, the hot regime,
the freeze/CO\textsubscript{2} infeasibility, and the noise-chatter regimes
in a single run, precisely where the architectural choices we compare are
predicted to differ. The outdoor temperature is bounded at a
$-5$\,\textdegree C floor on day~1 and at a $+20$\,\textdegree C ceiling on
day~3, with day~2 acting as a transition and day~4 as a summer cool-down with
a door-open event. Two unmeasured CO\textsubscript{2} bursts (one in each
regime) and one $UA$ spike are added at predefined times in the disturbance
scenario. The disturbance events of a
representative 24-hour window are summarised in Table~\ref{tab:scenario}:
two operator-visible (known) disturbances, two unmeasured disturbances, and
six operator-issued setpoint changes through the day. The full 96-hour
sequence repeats the same event grammar across the four days at the
outdoor-temperature levels described above.

\begin{table}[!ht]
\centering
\caption{Representative 24-hour window of the 4-day mixed-season scenario:
every event class the controllers face. The full 96-hour sequence repeats
this event grammar across the four days at the outdoor-temperature levels
described in Section~\ref{sec:scenario}.
Known disturbances are visible to the operator and to any feed-forward logic;
unmeasured disturbances are felt only through their effect on the controlled
variables ($C$, $T$). $u_0$ is the operator-defined desired fan setpoint that
enters the chain at the lowest priority (Section~\ref{sec:arc}); it is
nominally $50\,\%$ and is changed by the operator on a daily schedule.}
\label{tab:scenario}
\small
\begin{tabular}{l l p{3.6cm} p{4.9cm}}
\toprule
\textbf{Type} & \textbf{Channel} & \textbf{Profile / event} & \textbf{Purpose in the test} \\
\midrule
Known        & $T_{\text{out}}(t)$
             & piecewise-linear $-5\to-15\to+2\to-8$\,\textdegree C
             & exercises the cold regime; stresses heater and freeze protection \\[2pt]
Known        & $n_{\text{cows}}(t)$
             & two-step $80\to100\to80$ across milking
             & shifts metabolic CO\textsubscript{2} and heat load on a known schedule \\[2pt]
\midrule
Unmeasured   & $q_{\text{CO}_2}^{\text{ext}}(t)$
             & 2-hour CO\textsubscript{2} burst, 06:00--08:00 (bedding off-gassing)
             & forces ventilation to react to a CO\textsubscript{2} excursion with no advance warning \\[2pt]
Unmeasured   & $UA^{\text{ext}}(t)$
             & 10-min $UA$ spike at 14:00 (door propped open)
             & sudden heat-loss transient that must be rejected by feedback alone (\emph{Zoom~B}) \\[2pt]
\midrule
SP change    & $u_0$ (operator default fan)
             & $50\to40$\,\% at 03:00 (night mode, quieter fans)
             & lowest-priority chain input changed by the operator \\[2pt]
SP change    & $u_0$ (operator default fan)
             & $40\to55\to65$\,\% at 08:00 / 10:00 (morning / day)
             & two ventilation regime switches as the day starts \\[2pt]
SP change    & CO\textsubscript{2} SP of \texttt{cc\_1000}
             & $1\,000\to1\,200\to1\,000$\,ppm at 16:00 / 19:00
             & farmer relaxes then restores comfort SP around milking peak \\[2pt]
SP change    & $u_0$ (operator default fan)
             & $65\to50$\,\% at 21:00 (evening reset)
             & return to nominal default; tests bumpless transition \\
\bottomrule
\end{tabular}
\end{table}

The piecewise-linear $T_{\text{out}}$ profile is the dominant slow disturbance:
its lowest excursion to $-15$\,\textdegree C occurs in the early hours and drives
the architecture into the freeze-protection regime where heater and fan trade off
explicitly. The CO\textsubscript{2}
burst and the $UA$ spike are deliberately unannounced: they are precisely the
events where feed-forward fails and the closed-loop dynamics dominate.

\subsection{Control architectures}
\label{sec:cow-archs}

We implement and evaluate four controller architectures (Table~\ref{tab:archs})
on the same cow-barn plant, run over the same case study and scenarios with the
same plant imperfections and noise seed. Architecture~D is a separate ablation
that removes the decomposition entirely.

\subsubsection{Deriving the agent roster from the control-theory-designed chain}
\label{sec:cow-decomp}

The agent roster is derived directly from the ARC chain that Skogestad has
proposed for the cow-barn case~\cite{skogestad2000plantwide,skogestad2026arc}:
seven SIMC-tuned PI controllers, a four-layer MIN/MAX selector chain on $u_1$,
and a split-parallel arrangement at the heater.\footnote{We use the term
\emph{selector chain} (or \emph{series}), rather than \emph{cascade}, for the
MIN/MAX selector network. The selectors operate on the same time scale: every
controller in the network is tuned with the same $\tau_c$, and there is no
master-to-slave setpoint handoff between them. We reserve \emph{cascade} for a
genuine hierarchical arrangement in which a slow master loop sets the setpoint
of a fast slave loop, as in the relation between the supervisory orchestrator
and the regulatory layer. The selectors form a one-way priority chain: the last
selector in the chain is the most critical, and the others may be given up when
constraints conflict.} The chain is the input to the
multi-agent design, not something to design around. Two rules then exhaust
the agent roster.

The first equivalence: each PI in the chain corresponds to one operator agent.
The PI controller is the basic regulatory unit of the chain, and it
defines a natural unit of agent responsibility: one CV, one MV, one
setpoint, one role in the priority order, and one decision per control
sample. The multi-agent reformulation assigns one operator agent to each
PI's role in the chain. There are exactly as many operator agents as
PIs; on the cow case the roster is fixed by the ARC chain described in the
previous subsection (Architecture~A, Figure~\ref{fig:arc}): six constraint
defenders on the fan ($u_1$) plus one heater-loop PI on $u_2$, with priority
ordering MIN/MAX-MIN-MAX from operator default to safety overrides:
\begin{center}
\texttt{tc\_5}, \texttt{cc\_1000}, \texttt{tc\_20}, \texttt{tc\_0},
\texttt{cc\_3000}, \texttt{tc\_heater}.
\end{center}
What the agent \emph{does} have is operator-like \emph{context}: knowledge
of what a PI controller is and how it relates a measurement to an MV,
knowledge of the SIMC tuning rule (Eq.~\ref{eq:simc})~\cite{skogestad2003simc}
and the principle of back-calculation anti-windup~\cite{astrom-pid}, and
the agent's own
$(\text{CV}, \text{MV}, \text{SP}, \text{priority}, \text{selector kind})$
identity. Each control sample, the agent uses this context directly to
turn its measurement into a candidate MV value, exactly as a trained
operator at a control panel would. The agent owns nothing beyond its own context
and decision logic; it does not see the other agents' state, the other
constraints, or the structure of the chain above it. An alternative
realisation in which an agent's decision step is delegated to an LLM
that reads the same context is the position occupied by the LLM
operator architecture described in Section~\ref{sec:llm-mas}.

The second equivalence: the chain's interaction logic becomes one orchestrator agent.
The selector network, the split-parallel arrangement, and any override
path are structural elements that define how the PIs combine. Yet they
are part of the control architecture as firmly as the PIs themselves:
they encode the priority order in which constraints are honoured under
conflict, the operator's desired $u_0$, the hand-off between fan-driven
and heater-driven temperature control, and the freeze-protection guard.
This logic does not fit into any single operator agent (no operator owns
more than one constraint), but it has to live somewhere in an agent
system. The orchestrator agent receives the operator agents' proposals
every sample, applies the MIN/MAX selectors, the split-parallel rule,
and the override layer in the order prescribed by the ARC chain of
Architecture~A (Figure~\ref{fig:arc}),
and emits the actually-applied $(u_1, u_2)$. The orchestrator is the
agent-level encoding of the parts of the chain that are not PIs.

The orchestrator does not take on a PI's role; it represents the inter-PI logic.
In Architecture~A (the monolithic ARC) this logic is realised as
hard-coded MIN/MAX statements; in Architecture~B it is realised as a
configurable rule-based component that always runs the same priority
chain as ARC and adds dwell-time hysteresis, trend anticipation, and
deterministic parameter scheduling; in Architecture~C the same priority
chain is preserved deterministically and a Claude Opus 4.7 LLM-based
supervisor is added on top of it, taking the chain-state snapshot as
context and emitting bounded scalar tuning parameters (dwell time, trend
gains) every supervisory tick. Its inputs are the operator agents'
proposals (and, for Architecture~C, the chain-state snapshot fed to
the LLM supervisor); its outputs are the chain's prescribed $(u_1, u_2)$
commands. The orchestrator's arbitration logic is given by control theory,
not learned or guessed.

The multi-agent reformulation, though prescribed by control theory, adds value
in two ways. First, it makes the chain's state distributed and
auditable: every agent owns a small workspace and writes an append-only
log of its proposals, so any control decision can be replayed from the
filesystem alone (Section~\ref{sec:impl}). Second, it
makes the architecture pluggable along well-defined interfaces: agents
can be swapped between PI and rule-based variants without modifying the
orchestrator (Architectures~A--B), and the orchestrator can be swapped
between rule and smart variants without modifying the agents
(Architectures~B and C).

\
The agents in this paper fall into two categories that occupy different
tiers of the control hierarchy and play different roles relative to the
PI controllers of the chain:
\begin{itemize}[leftmargin=22pt,itemsep=2pt]
  \item Regulatory operator agents (\texttt{tc\_5},
        \texttt{cc\_1000}, \texttt{tc\_20}, \texttt{tc\_0},
        \texttt{cc\_3000}, \texttt{tc\_heater}) and the orchestrator
        agent. These run on the fast 2\,s control loop. Each operator
        agent acts as the equivalent of one PI in the chain: it
        follows the ARC principles and uses its control-theory context
        (CV, MV, setpoint, priority, selector kind, and the rolling
        memory of its own past proposals) to translate the current
        measurement into a proposal directly. The orchestrator agent
        encapsulates the chain's interaction logic (MIN/MAX selectors,
        split-parallel, overrides) and applies it to the operator agents'
        proposals.
\end{itemize}
Three architectures are constructed from the same ARC chain, each
serving as one ablation step in the evaluation of the proposal made in
this work.
Architecture~A is the classical monolithic baseline: six PI controllers
in a hard-coded MIN/MAX chain, no agents.
Architecture~B introduces the LLM operator agents while keeping the
orchestrator deterministic: each PI's role in the chain is taken by
one Qwen 2.5 7B Instruct operator agent, and the MIN/MAX chain is
encapsulated as one rule-based orchestrator agent that applies the
structural priority order without LLM involvement.
Architecture~C, the headline of this work, additionally delegates the
orchestrator-level reasoning to an LLM. The operator agents are the
same Qwen 2.5 7B Instruct ensemble used in Architecture~B; a Claude
Opus~4.7 supervisor classifies the operating regime and tunes the
orchestrator's dwell-time, trend-bias, and memory parameters at each
supervisory tick, while the structural MIN/MAX priority order remains
enforced deterministically. The three architectures share a common
agent roster (one operator agent per PI in the chain, one
orchestrator agent for the chain's interaction logic) and are
evaluated on the same disturbance scenarios.

\begin{table}[!ht]
\centering
\caption{The three evaluated architectures, derived from the same
ARC chain and forming a stepwise ablation: A is the monolithic
PI baseline; B replaces the PIs with Qwen 2.5 7B Instruct operator
agents while the orchestrator stays rule-based; C additionally
introduces a Claude-Opus~4.7 LLM at the orchestrator-tuning tier
while the structural priority chain is preserved in every case.}
\label{tab:archs}
\small
\begin{tabular}{c p{3.6cm} p{3.4cm} p{4.7cm}}
\toprule
\textbf{\#} & \textbf{Architecture} & \textbf{Operator} & \textbf{Orchestration} \\
\midrule
A & ARC (PI chain)                       & PI controllers          & rule-based MIN/MAX \\
B & LLM-MAS (Qwen 7B) + rule orchestrator  & Qwen 2.5 7B Instruct    & rule-based MIN/MAX \\
C & LLM-MAS (Qwen 7B) + LLM orchestrator   & Qwen 2.5 7B Instruct    & Claude Opus~4.7 supervises MIN/MAX \\
\midrule
D & Single-LLM (monolithic, ablation)      & one Qwen 2.5 7B (all constraints) & none: the model sets $u_1,u_2$ directly \\
\bottomrule
\end{tabular}
\end{table}

Architectures~A--C form a stepwise capability ladder built from the same ARC
chain; Architecture~D is a separate \emph{ablation} that removes the
decomposition itself. A single Qwen 2.5 7B Instruct model receives the entire
control problem (every measurement, every constraint and its priority, and
both manipulated variables) and must output $u_1$ and $u_2$ directly, with no
operator agents, no MIN/MAX chain and no orchestrator. It tests the
Introduction's claim that the ARC-style decomposition is what keeps each
agent's context narrow enough to be specified and verified; Architecture~D is
described in Section~\ref{sec:arch-d} and evaluated in
Section~\ref{sec:res-sigurd}.
Table~\ref{tab:translation} summarises the mapping between the
agent-level constructs used in this work and their established counterparts
in classical process control theory. Each row names a familiar
process-control object on the left, the agent-level counterpart used in the
present work in the middle, and the canonical reference on the right.

\begin{table}[!ht]
\centering
\caption{Translation grammar: every agent-level construct in this work has an
established counterpart in classical process control theory.}
\label{tab:translation}
\small
\begin{tabular}{p{6cm} p{6cm}}
\toprule
\textbf{Process-control construct} & \textbf{Agent-level counterpart in this work} \\
\midrule
PI controller targeting one constraint setpoint & Operator agent (one agent per PI in the chain) \\
MIN/MAX selector network & Orchestrator agent's priority-grouped decision rule \\
Back-calculation anti-windup & Bumpless agent re-engagement \\
Split-range / split-parallel & Two-MV decomposition (fan / heater) \\
Supervisory layer (gain scheduler) & LLM advisor on slow tier \\
Active-constraint switching & Priority-chain reordering by orchestrator \\
Plant-wide design procedure & Recipe for the agent roster (Section~\ref{sec:systematize}) \\
\bottomrule
\end{tabular}
\end{table}

\subsubsection{Systematizing multi-agent construction: a four-step recipe}
\label{sec:systematize}

The agent roster of Section~\ref{sec:cow-decomp} was derived from a
control-theory-designed chain. The same derivation
generalises to any plant for which a plant-wide regulatory design
exists, and produces a four-step recipe for constructing a multi-agent
regulatory layer:
\begin{enumerate}[leftmargin=22pt,itemsep=2pt]
  \item \emph{Design the regulatory layer with the plant-wide procedure}
        \cite{skogestad2000plantwide,skogestad2026arc}. Identify the
        controlled variables, the manipulated variables, and the active
        inequality constraints; determine where a constraint conflict
        requires CV--CV switching (MIN/MAX selectors) and where one
        manipulated variable must be shared by MV--MV switching
        (split-range or split-parallel); and add override paths for the
        safety constraints. The output of this step is the set of logical
        control blocks and the interaction logic that binds them.
  \item \emph{Assign one operator agent to each control loop.} The agent
        inherits the loop's external interface
        ($\texttt{measurement} \to \texttt{requested MV}$) and its place
        in the priority order, and acts as an operator carrying
        control-theory context rather than a numerical gain. There is one
        operator agent per loop, and the agents inherit the priority
        structure through their proposal tags.
  \item \emph{Encapsulate the interaction logic as one orchestrator agent.}
        The MIN/MAX selectors, the split-parallel, and any override paths
        are not control loops and do not fit inside any operator agent,
        yet they are part of the architecture. The orchestrator agent
        receives the operator agents' proposals every sample and emits the
        prescribed $(u_1, u_2)$ commands; it does not replace a loop, it
        represents the logic that binds them.
  \item \emph{(Optional) Add supervisory agents on top of the
        orchestrator.} The plant-wide hierarchy already separates
        regulatory control from supervisory control
        \cite{skogestad2000plantwide,engell2007}. In the agent system,
        this translates to additional agents that run on a slower
        time-scale and write their advice into bounded scalar parameters
        of the orchestrator, never into $u_1$ or $u_2$ directly. The
        smart-orchestrator's dwell-time and trend gains are an instance
        of this fourth tier and are exercised by Architecture~C.
\end{enumerate}
Steps 1--3 produce the basic agent system and account for
Architectures~A and B in this paper: A is the monolithic implementation
of the chain, B is the multi-agent reformulation under the recipe
above with rule-based operator agents and a rule-based orchestrator.
Architecture~C exercises step~4 by using LLM-backed operator agents
under the smart orchestrator. A validation swap in which the operator
agents' decision logic is implemented by SIMC-tuned PI blocks for
debugging purposes only reproduces ARC bit-for-bit, certifying that
steps~1 and 3 (the chain design and the orchestrator-agent encoding
of the interaction logic) are faithful.

The multi-agent reformulation rests on two structural equivalences:
\emph{each PI in the chain corresponds to one operator agent that uses
control-theory context to act directly on its proposal}, and \emph{the
chain's inter-PI logic becomes one orchestrator agent}. The reformulation
is extended by a prompt engineering and context curation step, which
replaces the PID tuning step of classical ARC.

\subsubsection{Architecture A: Advanced Regulatory Control (ARC)}
\label{sec:arc}

\begin{figure}[ht!]
\centering
\includegraphics[width=\linewidth]{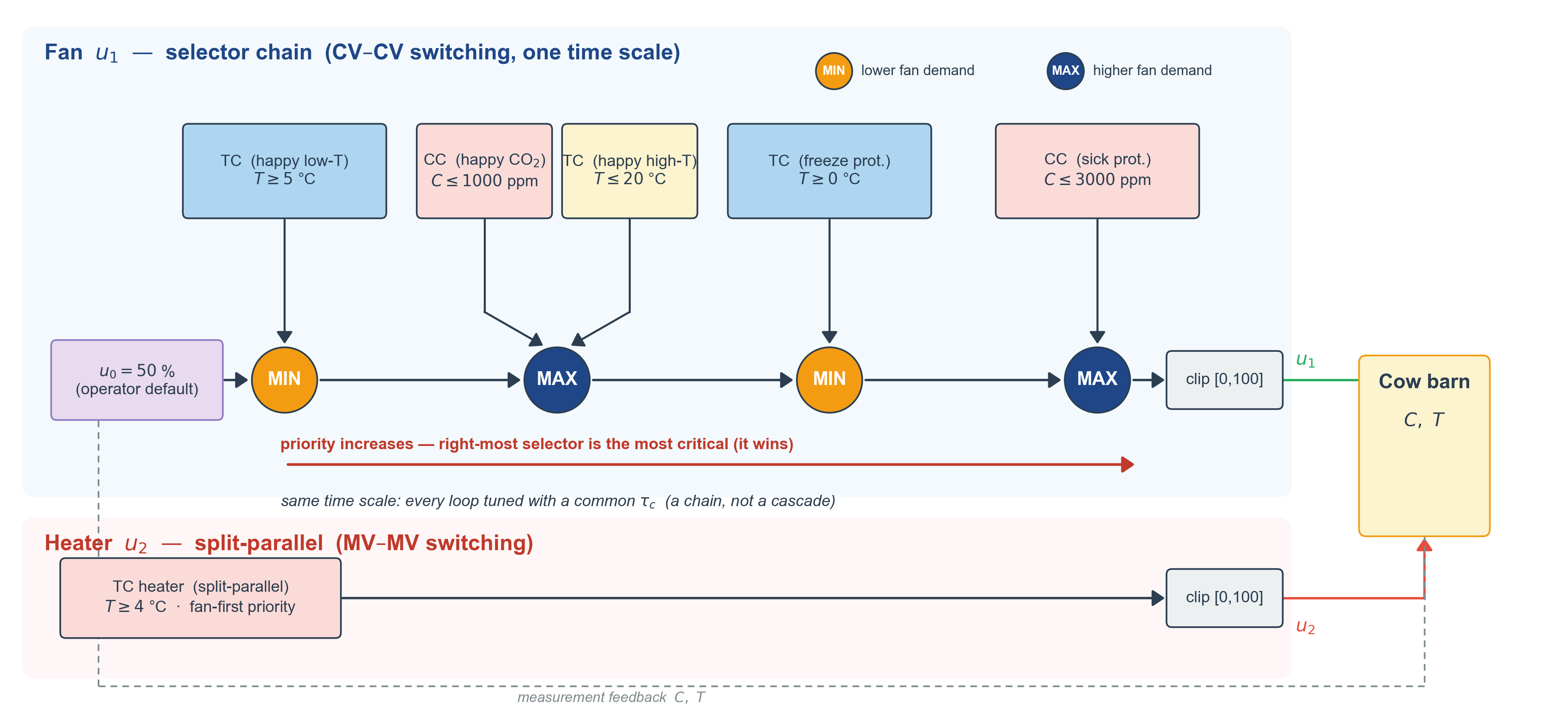}
\caption{Architecture A: Advanced Regulatory Control. Seven PI controllers,
a four-stage MIN/MAX selector chain on the fan ($u_1$), and a split-parallel
heater branch ($u_2$). The selector chain is drawn horizontally on purpose:
the four selectors form a one-way priority series on a single time scale
(every loop tuned with a common $\tau_c$, with no master-to-slave setpoint
hand-off), so it is a chain rather than a cascade. Priority increases from
left to right, and the right-most selector is the most critical: it overrides
the others when constraints conflict. CO\textsubscript{2} and the high-T
comfort target share the second-stage MAX. The heater is a separate
split-parallel branch (MV--MV switching, fan used before heater). Based on the
author's re-implementation following the architecture described
in~\cite{skogestad2026arc}.}
\label{fig:arc}
\end{figure}

This is the baseline: a faithful re-implementation of Figure~10 of
Skogestad~\cite{skogestad2026arc}. Six PI controllers each watch a single CV and
target a single constraint setpoint. Their output candidates are routed into a
four-layer MIN/MAX selector chain on $u_1$, while a seventh PI controls $u_2$
directly through a split-parallel arrangement~\cite{forsman2025splitparallel}
(Figure~\ref{fig:arc}).
Six PI controllers are assigned to the chain, each defending a single constraint
setpoint: \texttt{tc\_5} (T, SP=5, MIN); \texttt{cc\_1000} (C, SP=1000, MAX);
\texttt{tc\_20} (T, SP=20, MAX); \texttt{tc\_0} (T, SP=0, MIN, freeze override);
\texttt{cc\_3000} (C, SP=3000, MAX, sick-cow override); \texttt{tc\_heater}
(T, SP=$5-\Delta=4$\,\textdegree C, drives $u_2$).

The four-layer MIN/MAX selector chain on $u_1$, reading from lowest priority
(operator default) to highest (safety overrides), is implemented as follows:
\begin{lstlisting}[language=Python,numbers=none]
u <- u0 = 50%                                  # operator default
u <- MIN(u, tc_5.out)                          # happy low-T
u <- MAX(u, cc_1000.out, tc_20.out)            # happy CO2 / happy high-T
u <- MIN(u, tc_0.out)                          # freeze protection
u <- MAX(u, cc_3000.out)                       # sick-cow protection
u1 <- clip(u, 0, 100)
\end{lstlisting}
The split-parallel heater path is independent: $u_2 \leftarrow \mathrm{clip}(\text{tc\_heater output}, 0, 100)$.
The $\Delta = 1$\,\textdegree C offset between \texttt{tc\_5} and \texttt{tc\_heater} ensures
the cheap option (reducing fan to retain heat) is exhausted before the expensive option
(turning the heater on) is engaged.

The split-parallel arrangement adopted here differs from conventional split-range
control in a way that is not innocuous for the chain's clean operation:
the conventional MV--MV switch for temperature is \emph{split-range} control with
one setpoint at $T = 5$\,\textdegree C and the heater turned on by a
deadband on the negative-error half. We adopt the \emph{split-parallel}
arrangement of Forsman~\cite{forsman2025splitparallel} (two setpoints
separated by $\Delta = 1$\,\textdegree C) because it lets the fan and the
heater operate as cleanly decoupled SISO loops without a deadband-and-hysteresis
pair: the chain simply hands $u_1$ off via the MIN/MAX selectors, while
\texttt{tc\_heater} drives $u_2$ to its own slightly-lower setpoint. A pure
split-range implementation with a single $T = 5$\,\textdegree C target is a
defensible alternative that we do not evaluate here.

The chain above does not currently impose a backoff on the
\texttt{cc\_1000} or \texttt{tc\_5} setpoints. That is, the constraints
$C \le 1\,000$ and $T \ge 5$ are tracked at exactly $1\,000$\,ppm and
$5$\,\textdegree C. In practice the noise+actuator-lag dynamics produce
short-term overshoots, and the standard remedy is to tighten the operating SP
by a backoff $\Delta_C, \Delta_T$ so the constraint is honoured even at the
peak of the disturbance response~\cite{skogestad2026arc}.
$\Delta_C$ and $\Delta_T$ are tunable: an operator (or a supervisory agent)
can reduce them during quiet periods to save energy and increase them when
disturbances are expected. In the agent reformulation, this knob is the most
natural target for a strategy-level supervisor; we treat it as a planned
extension for future work.

\FloatBarrier

\subsubsection{PI controllers and SIMC tuning}

The SIMC procedure described below serves two purposes in this work. It
informs the gain selection for the PI controllers of Architecture~A, and it
also informs the operator agents of Architectures~B and C: the SIMC tuning
results, the binding direction, and the proportional-band interpretation are
all part of the context that each LLM operator agent receives in its prompt.
Each PI uses the {\AA}str{\"o}m-form
parallel-incremental algorithm:
\begin{equation}
u_{\text{raw}}(k) = K_c\,e(k) + I(k), \qquad e(k) = SP - y(k),
\label{eq:PI-raw}
\end{equation}
with the integrator updated by back-calculation anti-windup~\cite{astrom-pid}:
\begin{equation}
I(k+1) = I(k) + \Delta t \cdot \Bigl[\tfrac{K_c}{T_i}\,e(k) + \tfrac{1}{T_t}(u_{\text{act}}(k) - u_{\text{raw}}(k))\Bigr].
\label{eq:PI-aw}
\end{equation}
Here $u_{\text{act}}$ is the MV value \emph{actually applied} to the plant, which may
differ from $u_{\text{raw}}$ because of selector overrides or saturation, and
$T_t = T_i$ is the tracking time constant. This keeps overridden controllers'
integrators tracking reality so they re-engage cleanly when activated.

Gains are computed by Skogestad SIMC tuning~\cite{skogestad2003simc} of the
linearised plant about the nominal operating point. It is possible to verify from
Table~\ref{tab:simc} that the CO\textsubscript{2}--fan loop ($K_c = -0.173$\,\%/ppm)
requires a negative proportional gain: a positive error $e = SP - C$ means $C < SP$,
so $u_{\text{raw}} > 0$ requests less fan, which is correct for a MAX-type agent that
should request fan only when $C$ is above its setpoint. The temperature--fan loops
use the same sign convention with $K_c < 0$ for TC agents, while the heater loop uses
$K_c > 0$ because increasing $u_2$ raises $T$ directly. The three distinct loop gains
are summarised in Table~\ref{tab:simc}.

\begin{equation}
K_c = \frac{1}{K_p}\,\frac{\tau}{\tau_c + \theta}, \qquad
T_i = \min(\tau,\; 4(\tau_c + \theta)),
\label{eq:simc}
\end{equation}
The well-mixed plant model has no transport delay, but the $\tau_{\text{act}} = 20$\,s
first-order actuator lag (Section~\ref{sec:imperfections}) acts as an effective
delay $\theta_{\text{eff}} \approx 20$\,s as far as the controller is concerned. We
choose $\tau_c = 0.6\,\tau$ (mildly aggressive; closed-loop $\approx 1.7\times$
faster than open-loop) and equivalently set $\tau_c + \theta_{\text{eff}} = 0.6\,\tau$;
since $\theta_{\text{eff}} \ll \tau \approx 387$\,s, the resulting $K_c$ and $T_i$
agree with the $\theta = 0$ choice to all reported digits.

\begin{table}[!ht]
\centering
\caption{SIMC PI gains at the nominal operating point ($u_1=50\,\%$, $n_{\text{cows}}=80$).}
\label{tab:simc}
\begin{tabular}{l c c c c}
\toprule
\textbf{Loop} & $K_p$ & $\tau$ [s] & $K_c$ & $T_i$ [s] \\
\midrule
CO\textsubscript{2}\,$\leftrightarrow$\,fan (\texttt{cc\_*})  & $-9.66$\,ppm/\%  & 387 & $-0.173$\,\%/ppm & 387 \\
T\,$\leftrightarrow$\,fan (\texttt{tc\_5/0/20})               & $-0.119$\,K/\%   & 334 & $-14.02$\,\%/K   & 334 \\
T\,$\leftrightarrow$\,heater (\texttt{tc\_heater})            & $+0.092$\,K/\%   & 334 & $+18.08$\,\%/K   & 334 \\
\bottomrule
\end{tabular}
\end{table}

All PIs use the same $\Delta t = 2$\,s sample period.

\subsubsection{Architectures B and C: LLM operator agents on the regulatory tier}
\label{sec:llm-mas}

Architecture~A keeps the operator agent's decision logic deterministic
as a SIMC-tuned PI block. Architectures~B and~C share a common
implementation of the LLM operator agents that replaces this deterministic decision
logic with a local instruction-tuned language model: each of the six
operator agents \texttt{tc\_5}, \texttt{cc\_1000}, \texttt{tc\_20},
\texttt{tc\_0}, \texttt{cc\_3000}, \texttt{tc\_heater} is backed by
Qwen~2.5 7B Instruct (7.62\,B parameters, FP16, served by a single
shared inference backend running offline on a 24\,GB consumer GPU). The
two architectures differ only at the orchestrator tier: Architecture~B
uses a deterministic rule-based orchestrator that applies the MIN/MAX
priority chain exactly as described in Section~\ref{sec:arc} and
makes no further reasoning step; Architecture~C adds a Claude Opus~4.7
supervisor that classifies the operating regime and tunes the
orchestrator's dwell-time, trend-bias and memory parameters at each
supervisory tick. In both cases the structural MIN/MAX priority chain
is preserved deterministically and the LLM cannot override the chain's
binding-direction decisions; the operator-tier LLM contributes a regime
classification and a rationale, the orchestrator-tier LLM (when present)
contributes parameter tuning, and the chain enforces structural
priority regardless of either output. This is the architectural
realisation of conflict resolution by structural priority that motivates
the paper.

The construction of Architectures~B and~C is fully prescribed by the
four-step recipe of Section~\ref{sec:systematize}: design the chain
with the plant-wide procedure, assign one operator agent to each PI's
role, encapsulate the inter-PI logic as one orchestrator agent, and
(optionally) add an LLM-based supervisory agent on a slow tier.
Architecture~B exercises the first three steps; Architecture~C
additionally exercises the fourth.

\
Architecture~C runs at $\Delta t_\text{ctrl} = 5$\,min, an order of
magnitude slower than the regulatory tier of Architectures~A--D
($\Delta t = 2$\,s).  This is the natural cadence of an operator
glancing at a SCADA panel rather than the cadence of a fast PI loop.
The plant time constants are $\tau_C \approx 387$\,s and
$\tau_T \approx 334$\,s, so $\Delta t / \tau \approx 0.83$; the
plant is partially settled between samples but not fully (which would
otherwise drive the chain into bang-bang behaviour).  The plant
itself integrates continuously between samples through a stiff ODE
solver; only the controller's polling rate slows.

We chose Qwen 2.5 7B Instruct as the headline backend for four reasons.
\emph{(i) Instruction-following capacity.}  The agent's task is to
classify the regime (\texttt{IDLE}~/~\texttt{ACTIVE}~/~\texttt{SATURATED})
from a pre-formatted state report and return a JSON object with a
short operator-style rationale.  At 7\,B parameters Qwen 2.5 follows
this format reliably: across the 4-day run reported in
Section~\ref{sec:res-llm-mas}, $0$ of the LLM calls failed
schema validation.  A 3\,B-parameter version of the same family
(Qwen 2.5 3B Instruct) was also exercised in development; it
satisfied the schema but produced wrong-direction proposals on
$\sim$60\,\% of saturated MAX agents in an earlier prompt design,
behaviour we explicitly engineered out via the deterministic-direction
mapping described below (Lever~A). A plausible hypothesis for the
3B-model failure is that the instruction-tuning corpus carries little
process-control vocabulary, so the model pattern-matches phrases such
as ``high CO\textsubscript{2}'' to a general ``reduce activity'' response
rather than to the loop-specific ``increase ventilation'' action, and
the 3-billion-parameter ceiling appears to be insufficient for the
multi-constraint long-context reasoning that the saturated-MAX case
requires. The 7B variant retains the same general-purpose
instruction-tuning bias but is capable of integrating the explicit
binding-direction line of the system prompt without inverting it, which
is consistent with the $0\,\%$ schema-validation failure rate observed on
the 4-day run.
\emph{(ii) Open weights, offline-deployable.}  Qwen 2.5 is released
under the Apache-2.0 licence; the model weights are downloaded once
($\sim 15$\,GB on disk in FP16) and the entire run is offline.  This
matters for industrial process control, where cloud connectivity may be
restricted, where data-access constraints may forbid the transmission of
operating-point data outside the plant boundary, and where security and
intellectual-property concerns require that the inference layer remain
on the plant network rather than on a third-party cloud. It also matters
for paper reproducibility: the model
weights are pinned, no API drift is possible, and one with the
same GPU class is capable of reproducing every $(C, T, u_1, u_2)$ trajectory of this
section by cloning the repository and downloading the same checkpoint.
\emph{(iii) Single consumer GPU.}  In FP16 the model fits in
$\sim$15\,GB of VRAM with $\sim$9\,GB of headroom for the shared
system-prompt KV cache and the per-sample suffix processing, well
within a 24\,GB-class GPU.
\emph{(iv) Marginal cost is electricity.}  No API spend; the entire
4-day mixed-season scenario completes in approximately one hour of GPU time.

\
A useful framing of the LLM operator agent, contrasted with the classical
PI block of Architecture~A, is that the LLM agent does not require gain
tuning in the SIMC sense. What it requires instead is to be
\emph{educated}: a prompt-engineering step that fixes the agent's role
and a context-curation step that supplies the loop-specific reference
material the agent reasons against. The prompt engineering replaces the
SIMC tuning step of the classical pipeline, and is itself the human
design effort that turns a general-purpose instruction-tuned model into
a specialised operator agent. Each LLM operator agent receives a
three-layer prompt at every control sample: a static \emph{system prompt}
shared across all six agents ($\sim$1\,400 tokens, KV-cached), a
per-agent \emph{identity prompt} ($\sim$80 tokens, fixed for one agent
across a scenario), and a per-sample \emph{user prompt} ($\sim$140
tokens, refreshed each control tick). The three layers together limit the agent's context to
its own constraint, setpoint, chain priority, the last few
measurements, and its own past mode classifications; the agent does
not see other agents' state and does not see the chain structure
above it. This is the engineering-handcraft alternative to RAG that
the Introduction motivates: the specialisation problem is solved once,
at design time, by encoding the agent's role in its system prompt, and
the context required by any one agent is limited by construction.

The system prompt is reproduced verbatim below. It is the
agent's operator-style training material, distilled from Skogestad's
2023 review of advanced regulatory
control~\cite{skogestad2023arc}, which is included verbatim in the
operator agent's context as the canonical reference for the chain
vocabulary and the selector rules the agent is part of. The two
foundational selector rules from that review are quoted as canonical
references inside the prompt so the LLM understands the chain it is
part of, but the selector vocabulary is reserved for the orchestrator.

\begin{lstlisting}[language=,basicstyle=\ttfamily\footnotesize,numbers=none,breaklines=true,frame=single,framerule=0.3pt]
You are an operator agent in a multi-agent regulatory control system for a
dairy barn. The architecture follows the Advanced Regulatory Control (ARC)
pattern of Skogestad (Annual Reviews in Control, 2023).

You and five peer agents each defend ONE inequality constraint on the plant.
A separate Orchestrator agent collects your proposals and applies the
priority chain described below; you do not need to know the chain rules
to do your job. You only need to defend YOUR own constraint.

PLANT
  C  = CO2 concentration in the barn [ppm]
  T  = barn air temperature [degC]
  u1 = fan speed [0..100 %]    (increasing u1 LOWERS both C and T)
  u2 = heater [0..100 %]       (increasing u2 RAISES T)

CONSTRAINT-CHASING DIRECTIONS  (your identity tells you which is yours)
  Each constraint is "satisfied by a LARGE MV" or "satisfied by a SMALL MV".
  This is Skogestad 2023 Selector Rule 1, which the chain uses to decide
  MIN vs MAX selectors. For YOU as an operator the meaning is simple:
    "satisfied by LARGE MV"  ==  when violated, you push the MV UP.
    "satisfied by SMALL MV"  ==  when violated, you push the MV DOWN.
  In both cases your identity provides two concrete numbers:
    u_idle      = the value you propose when your constraint is comfortably
                  satisfied (NON-binding; lets other agents win the chain).
    u_saturate  = the value you propose when your constraint is clearly
                  violated (FULL bind).

YOUR JOB EACH CONTROL SAMPLE
  You receive (in the user prompt):
    - the current measurement of your CV
    - your setpoint SP
    - the signed violation v = (unsafe-side excess), positive = violated
    - the half-width of your proportional band PB/2
    - a pre-computed mode hint based on |v| vs PB/2

  Decide which of three regimes you are in:
    IDLE       v < -PB/2          your CV is comfortably on the safe side
    ACTIVE     -PB/2 <= v <= PB/2 your CV is inside the band; bind proportionally
    SATURATED  v >  +PB/2         your CV has clearly violated the constraint

  Then write a short operator-voice rationale (one sentence, <= 80 chars).
  THE NUMERIC u_request IS COMPUTED BY THE AGENT CODE FROM YOUR MODE.

OPERATOR MEMORY (will appear in your user prompt as a "MEMORY" block)
  You also have memory of how long the constraint has been violated and
  the integrated drift since the last comfortably-safe interval. Use this
  when picking ACTIVE intensity: if u_extra is already large, pick a
  moderate intensity rather than 1.0 to avoid double-strength action.

SITUATIONAL AWARENESS ("RECENT HISTORY" block)
  You also see your own last few mode classifications and the resulting
  CV trajectory. If you have been flipping IDLE <-> SATURATED rapidly
  and the resulting CV peak-to-peak is large, your decisions are causing
  bang-bang. When the user prompt flags oscillating: true, prefer ACTIVE
  with a moderate intensity over the recommended mode.

OUTPUT FORMAT (return EXACTLY one JSON object):
  {"mode": "IDLE|ACTIVE|SATURATED",
   "intensity": <float 0..1, used only for ACTIVE>,
   "rationale": "<one short operator sentence, <= 80 chars>"}

\end{lstlisting}

The per-agent identity prompt for
\texttt{cc\_1000} (the CO\textsubscript{2} comfort agent) is:

\begin{lstlisting}[language=,basicstyle=\ttfamily\small,numbers=none,breaklines=true,frame=single,framerule=0.3pt]
YOUR IDENTITY
  name      = cc_1000
  CV        = C
  MV        = fan
  setpoint  = 1000 ppm
  constraint: C <= 1000 ppm
  binding direction: push MV UP when violated
  proportional band PB = 100 ppm
  u_idle      = 0 %  (when not binding)
  u_saturate  = 100 %  (when fully binding)

You only emit {mode, intensity, rationale}; the agent code
will set u_request from your mode and the two values above.
\end{lstlisting}

A representative per-sample user prompt for \texttt{cc\_1000} at $t = 1200$\,s with
the constraint approaching its limit reads:

\begin{lstlisting}[language=,basicstyle=\ttfamily\small,numbers=none,breaklines=true,frame=single,framerule=0.3pt]
CURRENT STATE  (t = 1200 s)
  measurement = 985.0 ppm
  setpoint    = 1000 ppm
  v           = -15.00 ppm (positive = unsafe side), trend = +2.30 ppm/min
  zone        : |v|=15.0 <= PB/2=50.0, inside band (linear alpha = 0.35)
  recommended mode (from rule)  : ACTIVE
  previous u_request  = 12.0 %

MEMORY (your operator log; updates every sample):
  binding duration : 0 s continuous on the unsafe side
  integrated drift : 0.000 ppm*s
  u_extra applied  : +0.00 %  (already added; Ti=600 s)

RECENT HISTORY  (your own last 12 samples = ~60 min):
  your modes (oldest -> newest, I=IDLE A=ACTIVE S=SATURATED):
    IIIIIIIIIAAA
  resulting C measurements: [850.0, 860.0, 870.0, 880.0, 890.0, 910.0,
                              930.0, 950.0, 960.0, 970.0, 975.0, 985.0] ppm
  peak-to-peak: 135.00 ppm     mode flips: 2

Return one JSON object (mode, intensity, rationale).
You are STRONGLY encouraged to follow the recommended mode.
\end{lstlisting}

Three elements of the user prompt warrant emphasis. The
\emph{recommended mode} is a deterministic hint computed from the
signed violation $v$ and the proportional band $\mathrm{PB}/2$; the LLM
is encouraged to follow it but may override it on the basis of trend
or recent history. The \emph{memory block} surfaces the agent's
integral state $u_{\text{extra}} = K_i \cdot e_{\text{int}}$, already
applied by the agent code (Lever~B, Eq.~\ref{eq:lever-B-u}); the LLM
uses this to avoid double-counting accumulated stress when choosing
intensity. The \emph{recent history block} carries the last twelve
mode classifications and CV measurements (Lever~D); when the
mode-flip rate exceeds 40\,\% of the window, the block is annotated
\texttt{oscillating: true} and the LLM is explicitly authorised to
break the recommended-mode default and stabilise on a moderate ACTIVE
intensity.

Each LLM call returns exactly one JSON object with three fields:
\begin{lstlisting}[language=Python,numbers=none]
{"mode":      "IDLE | ACTIVE | SATURATED",
 "intensity": <float 0..1, used only for ACTIVE>,
 "rationale": "<one short operator sentence, <= 80 chars>"}
\end{lstlisting}
The agent code then computes $u_\text{request}$ \emph{deterministically}
from the agent's identity:
\begin{equation}
u_\text{request} =
\begin{cases}
u_\text{idle} & \text{if mode = IDLE} \\
u_\text{saturate} & \text{if mode = SATURATED} \\
u_\text{idle} + \text{intensity} \cdot (u_\text{saturate} - u_\text{idle})
& \text{if mode = ACTIVE.}
\end{cases}
\label{eq:lever-A}
\end{equation}
The result is post-clamped to the slew-rate budget
$|u - u_\text{prev}| \le \text{rate}_\text{max}\,\Delta t$ and to
$[0, 100]$\%.

This mapping is the architectural fix we call \emph{Lever A}. The reasoning
behind it follows the storytelling of how the prompt evolved across earlier
trials, and is presented here because the design choices are easier to
follow when the failure modes that motivated them are visible. The first
prompt design asked the LLM to emit $u_\text{request}$ directly, as a
numerical fan or heater value between 0 and 100. The result was a high rate
of direction-inversion failures, around 60\,\% on saturated MAX agents under
stress: the model pattern-matched ``high CO\textsubscript{2}'' to ``shut
something down'' or ``reduce the action'', even when the correct numerical
value was listed in the prompt and the binding direction was stated in
words. The model was being asked to do two things at once: read the regime
and produce a number consistent with the binding direction, and it was
making the second decision incorrectly. Lever~A separates those tasks. The
binding direction is encoded in the pair $(u_\text{idle}, u_\text{saturate})$,
which is fixed in the agent's identity, not in the JSON the LLM returns.
The LLM cannot get the binding direction wrong by construction: if it says
SATURATED on a cc\_1000 agent with $u_\text{idle}=0,\ u_\text{saturate}=100$,
the resulting $u_\text{request}$ is $100$ regardless of any wrong-direction
language in the rationale. Lever~A removes this failure mode structurally;
the LLM contributes regime classification and an operator-voice rationale,
and direction is the agent's identity, not the LLM's choice. The same
storytelling motivates Levers B and D: each lever was added in response to
a specific failure mode observed in development. Lever~B was added after
trials in which the agent reached a steady-state offset on long disturbance
episodes and could not settle at an intermediate $u$. Lever~D was added
after trials in which the agent flipped between IDLE and SATURATED every
sample with no awareness of its own oscillation history.

A pure proportional band like Eq.~\ref{eq:lever-A} reaches a
steady-state offset on any sustained disturbance whose recovery
requires an intermediate $u$: the LLM's mode toggles between IDLE and
SATURATED but the agent has no memory of how long the constraint has
been violated, so it cannot settle at an intermediate $u$ between
$u_\text{idle}$ and $u_\text{saturate}$.  This is the same offset that
the I term of a PI controller eliminates by integrating the tracking
error.  In our case study it would otherwise produce bang-bang freeze
excursions when the heater alone cannot keep
$T \ge 5$~\textdegree C during a sustained $-10$~\textdegree C cold
day in the 4-day mixed-season scenario: the agent oscillates
between fan-off (saturate) and fan-on (\texttt{cc\_1000} winning the
chain because CO\textsubscript{2} climbs while the fan is off), and
the small steady $u$ that would settle $T$ near the constraint
boundary is never explored.

We close this gap with a second architectural lever, \emph{Lever~B}: an
\emph{operator memory} term carried by each LLM operator agent.  Each
agent maintains a standard integral of the signed violation $v$
(positive on the unsafe side), clipped at zero so it cannot push the
manipulated variable in the wrong direction:
\begin{equation}
e_\text{int}(k{+}1) =
\begin{cases}
\max\!\bigl(0,\, e_\text{int}(k) + \Delta t\,v(k)\bigr)
   & \text{if not at saturation,} \\
e_\text{int}(k)
   & \text{if } v(k) > 0 \text{ and } u_\text{prev} \text{ at saturate.}
\end{cases}
\label{eq:lever-B-eint}
\end{equation}
The integrator contributes a small extra term that is added to the
deterministic mode mapping of Eq.~\ref{eq:lever-A} before the slew and
clip stages:
\begin{equation}
u_\text{request} = \operatorname{clip}_{[0,100]}\!\!\Bigl(
   u_\text{request}^{\text{(Lever A)}}
   \;+\; K_i \cdot e_\text{int} \Bigr),
\quad
K_i = \frac{u_\text{saturate} - u_\text{idle}}{\mathrm{PB}\cdot T_i}.
\label{eq:lever-B-u}
\end{equation}
The integral gain $K_i$ inherits the binding direction from the
agent's identity (sign of $u_\text{saturate} - u_\text{idle}$); the
agent code does not pick a direction.  $T_i$ is an integral time
constant in the spirit of SIMC tuning~\cite{skogestad2003simc}: $T_i =
30$\,min for the temperature-loop agents (\texttt{tc\_5},
\texttt{tc\_20}, \texttt{tc\_heater}), $T_i = 15$\,min for the
freeze-protection agent \texttt{tc\_0}, and $T_i = 10$\,min for the
faster CO\textsubscript{2}-loop agents (\texttt{cc\_1000},
\texttt{cc\_3000}).

Equation~\ref{eq:lever-B-eint} is a standard PI integral with one
modification: the $\max(0, \cdot)$ clip ensures $e_\text{int}\ge 0$ at
all times.  This is the right choice for one-sided ARC constraints
like ``$T \ge 5$~\textdegree C'': the integrator accumulates only on
the unsafe side, and on the safe side ($v < 0$) it bleeds back to
zero \emph{linearly} in proportion to how safe we are, then locks at
zero.  This matches operator intuition. An operator accumulates
stress about a constraint going \emph{wrong}, not about it going
extra-right, and is also the right controls choice: a
bidirectional integrator would build up ``safe-side debt'' that
sabotages the next violation episode.  Anti-windup is conditional
integration: when $u_\text{prev}$ is already at the saturate end and
$v$ is still positive, $e_\text{int}$ is held; we also hard-clip
$|K_i \cdot e_\text{int}| \le 100$\,\% so the integral term alone
cannot demand more than full range.

The integrator state is surfaced to the LLM in the user prompt as a
\textsc{memory} block listing the binding duration, the integrated
drift, and the current $u_\text{extra} = K_i \cdot e_\text{int}$.  The
LLM is told that $u_\text{extra}$ has \emph{already} been added to its
mode-mapped output and that it should choose intensity for ACTIVE mode
taking this into account; this prevents the LLM from double-counting
accumulated stress when picking intensity.  Together with Lever~A,
this gives the LLM operator agent the same structural role as a PI
controller, with the operating point selected by reasoning
rather than by $K_c \cdot e$, and with the operator-style rationale
preserved as an audit-trail output.

Lever~A and Lever~B together address two of the three failure modes
of an LLM operator at the regulatory tier: direction-of-binding
(Lever~A) and slow-frequency steady-state offset (Lever~B). A third
failure mode, high-frequency bang-bang at the discrete-mode
classification boundary, arises when the LLM is asked to make a
fresh classification every sample with no awareness of its own past
decisions or their cumulative effect on the controlled variable.  An
operator looking at a SCADA trend chart for an hour would notice ``$T$
has been flipping between $-5$ and $+15$\,\textdegree C and I have
been flipping mode every other sample''; the agent as specified above
has no such situational awareness.  Lever~D supplies it.  Each
operator agent maintains a rolling window of its last $N$ mode
classifications and the corresponding CV measurements (we use $N =
12$, so the window covers the last hour at $\Delta t = 5$\,min), and
the user prompt includes a \textsc{recent history} block listing the
mode sequence (compact codes \texttt{I}/\texttt{A}/\texttt{S}), the
measurement values, the peak-to-peak amplitude, and the count of mode
flips.  When the mode-flip rate exceeds 40\,\% of the window length
($\ge 5$ flips in $12$ samples), the block is annotated
\texttt{<<- OSCILLATING}, and the system prompt explicitly authorises
the LLM to override the rule-derived recommended mode: pick
\texttt{ACTIVE} with a moderate intensity ($\sim 0.5$) and stay there
for several samples to let the plant settle into an intermediate
steady state.  This is operator situational awareness as a prompt
mechanism: the LLM is given enough long-horizon context to recognise
that its own decisions have been causing oscillation, and is given
explicit permission to break the recommended-mode default in response.
Lever~D is purely a prompt-engineering lever (it does not modify the
agent's control law) and is therefore an empirical test of whether a
7\,B-class instruction-tuned model can use long-horizon context to
self-regulate at the regulatory tier.

Each control sample, each agent appends one JSON record containing the
timestamp, measurement, setpoint, $u_\text{request}$, mode, the LLM's
rationale, and a fallback flag.  Across the 4-day mixed-season scenario reported
below this produces 6 time-aligned chains of operator rationales,
1\,152 entries each: a process-engineer's logbook of \emph{why}
each operator proposal was issued, not just what it was.  This is the
operational contribution of the LLM-as-operator pattern that the
deterministic Architecture~A cannot match without a separate
explanation layer.

The LLM call sits behind five layers of safety: (i) JSON schema
validation; (ii) a clip-and-slew clamp on $u_\text{request}$ to
$[0, 100]$\,\% with $|u - u_\text{prev}| \le \text{rate}_\text{max}
\Delta t$; (iii) Lever~A's deterministic-direction mapping
(Eq.~\ref{eq:lever-A}); (iv) Lever~B's anti-windup;
conditional integration freezes $e_\text{int}$ when the actuator is
saturated and the constraint is still violated, plus the hard clip
$|K_i \cdot e_\text{int}| \le 100$\,\% so the integral term alone
cannot demand more than full range; (v) a fallthrough to the rule
operator's proportional-band proposal (with $u_\text{extra}$ still
applied) on any LLM exception, schema violation, or JSON-parse
failure.  Empirically the fallthrough is rarely exercised (0
fallthroughs in the 4-day run reported below); the deterministic
fallback exists so a misbehaving model cannot leave the chain
without a proposal.

\

The two architectures differ only in the orchestrator. Architecture~B
uses the deterministic rule orchestrator described above: at every
control sample it groups the operator agents' proposals by chain
priority and applies the MIN or MAX selector at each priority level in
sequence, producing the two manipulated-variable commands
$(u_1, u_2)$ without any LLM call. The orchestrator's context is
limited by construction to the current sample's proposals plus the
agents' static identities; the orchestrator does not reason about the
operating regime and does not vary its parameters during a scenario.

Architecture~C introduces an LLM-based supervisor at the
orchestrator tier. A Claude Opus~4.7 advisor is called every
$\text{llm\_period}_s = 600$\,s (every 10 minutes of simulated time)
on a snapshot of the chain state. The advisor classifies the
current operating regime and returns three scalar tuning parameters
(a dwell time and two trend gains) that
the orchestrator uses for the next 10-minute window. The advisor's
context is limited to the slow tier: it cannot set $u_1$ or $u_2$ and
its outputs are bounded scalars validated against a fixed schema
before they are accepted by the orchestrator. The structural MIN/MAX
priority chain remains in effect throughout; the freeze-safety
guard at $T < 3$\,\textdegree C zeroes out the MAX-fan trend push
regardless of the advisor's recommendation.

The Claude advisor's verbatim system prompt, the canonical reference for the
LLM-based orchestrator's context, is reproduced below:

\begin{lstlisting}[language=,basicstyle=\ttfamily\footnotesize,numbers=none,breaklines=true,frame=single,framerule=0.3pt]
You are an expert supervisor for a multi-agent process control system
running inside a Norwegian dairy cow barn. Your task is to classify the
current operating regime and recommend tuning parameters for the smart
orchestrator.

SYSTEM CONTEXT
==============
The barn is controlled by six operator agents plus one orchestrator.
Operators propose requests for the fan (u1) or heater (u2); the
orchestrator resolves conflicts using a MIN/MAX selector chain by
priority, where higher priority is closer to the MV (more override
authority).

Agents (from highest priority to lowest):
  Priority 4  cc_3000   MAX   CO2 <= 3000 ppm    sick-cow safety
  Priority 3  tc_0      MIN   T   >= 0   C       freeze protection
  Priority 2  cc_1000   MAX   CO2 <= 1000 ppm    happy CO2
  Priority 2  tc_20     MAX   T   <= 20  C       happy high temperature
  Priority 1  tc_5      MIN   T   >= 5   C       happy low temperature
  Priority 0  tc_heater (direct)  split-parallel heater SP=4 C

The smart orchestrator has three knobs you adjust:

  dwell_s    : seconds an active-bind agent holds control after its raw
               proposal stops binding. Larger = smoother under noise,
               slower to react. Range [10, 600].
  k_trend_T  : gain on dT/dt biasing temperature proposals toward an
               approaching constraint. Range [0, 500].
  k_trend_C  : same for CO2 proposals. Subject to a hard safety guard:
               when T < 3 C, the orchestrator zeroes out the MAX-fan
               trend push regardless of your recommendation.

REGIME TAXONOMY
===============
Classify the current snapshot into exactly one of:
  nominal          - CVs comfortable, no active binding
  cold regime     - T dropping or below 5 C, heater saturated
  hot-saturation   - T rising near 20 C, fan at or near 100 %
  noise-chatter    - both CVs near a selector boundary, flipping
  sick-risk        - CO2 above 1500 ppm and climbing
  transient        - fast transition; keep moderate defaults

OUTPUT (strict JSON, validated by schema)
========================================
Return exactly:
  mode:       one of the six regimes above
  dwell_s:    float in [10, 600]
  k_trend_T:  float in [0, 500]
  k_trend_C:  float in [0, 500]
  rationale:  one short sentence explaining the choice (<=240 chars)

You have no access to u1 or u2. You cannot set MVs. Your output only
tunes the orchestrator's reasoning heuristics. The numerical MIN/MAX
chain retains final authority, and the freeze-safety guard is always
on regardless of your advice.
\end{lstlisting}

The supervisor's user prompt at each tick is a compact snapshot of the
chain state: simulated time, the two CV measurements with their
derivatives, the currently active priority level, the current regime
classification, and a per-agent table of proposals (CV, MV,
$u_\text{request}$, selector kind, priority). The output is constrained
to a fixed schema (a mode label, a dwell time, two trend gains, and a
rationale string) and the
call uses prompt caching for the static system prompt. The supervisor
sees only the chain state; it never sees the LLM operator agents'
rationales.

\begin{sidewaysfigure}
\centering
\includegraphics[width=0.95\textheight, keepaspectratio]{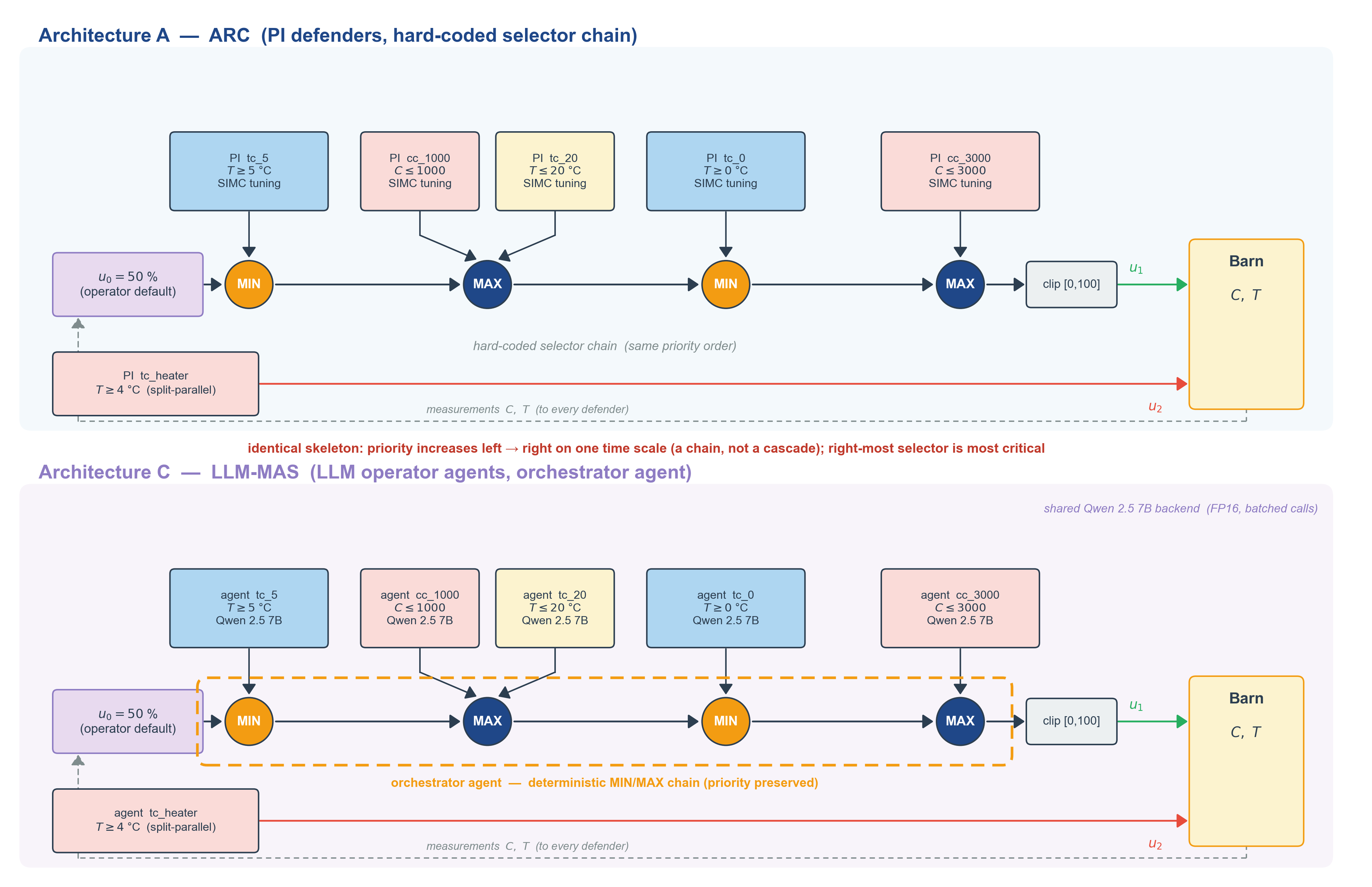}
\caption{Structural twins: Architecture~A (top, ARC) and Architecture~C
(bottom, LLM-MAS).  Both share an identical structural skeleton:
plant, six constraint-defenders, the four-stage MIN/MAX selector
chain drawn horizontally as a one-way priority series on a single
time scale (priority increasing left to right, right-most selector
most critical), the split-parallel heater branch, the clip and the
plant feedback.  The only differences are inside the
constraint-defender boxes (PI controller versus LLM operator agent
calling a shared Qwen 2.5 7B backend) and the chain implementation
(hard-coded code in A versus the orchestrator agent in C).  The
orchestrator agent is highlighted in the bottom panel with a dashed
amber boundary.}
\label{fig:arc-vs-llmmas}
\end{sidewaysfigure}

The two architectures are designed so the only difference is the
orchestrator-level reasoning. Architecture~B has a single LLM in the
loop, at the operator tier, and the orchestrator applies the priority
chain deterministically. Architecture~C adds a second LLM at the
supervisory tier; the structural priority chain is still enforced
deterministically by the orchestrator at the regulatory tier, and the
supervisor's role is to slowly retune the orchestrator's dwell-time
and trend-bias parameters in response to the operating regime. The
multi-tier LLM architecture preserves conflict resolution by
structural priority at the regulatory tier and uses LLM reasoning only
for parameter scheduling at the supervisory tier.
Figure~\ref{fig:arc-vs-llmmas} places Architectures~A and~C side by side:
they share an identical structural skeleton and differ only inside the
constraint-defender boxes and in how the priority chain is implemented.

\FloatBarrier

\subsubsection{Architecture D: monolithic single-LLM controller (ablation)}
\label{sec:arch-d}

Architectures~A--C all rest on the same decomposition: the plant-wide problem
is split into six narrow operator roles, each defending a single constraint
with a single manipulated variable and requiring only its own measurement,
setpoint and chain priority (Section~\ref{sec:cow-decomp}). The Introduction
argues that this decomposition is precisely what keeps each agent's context
narrow enough to be specified in a short, verifiable prompt. Architecture~D is
the ablation that tests that argument directly by removing the decomposition.

At each control step a single Qwen 2.5 7B Instruct model receives, in one
context, both measurements ($C$ and $T$), every constraint and its priority,
both manipulated variables, and a short rolling history; it must return the
fan command $u_1$ and the heater command $u_2$ directly. There is no operator
agent, no MIN/MAX selector chain and no orchestrator: the single model alone
must perform the multi-constraint priority coordination that the chain
performs structurally in A--C. Everything else is held identical: the same
plant, parameters, disturbance scenario, noise seed, $60$\,s control cadence,
and the same MV move-suppression limit applied to $u_1$ and $u_2$, so that any
difference in closed-loop behaviour is attributable to the loss of
decomposition alone, and not to a change in actuator authority, sampling rate
or measurement information. The monolith is given exactly the information the
operator team collectively holds (the two measurements, the operator's
comfort CO\textsubscript{2} setpoint and the default fan level) and no more;
in particular it is not shown the outdoor temperature or the cow count, which
the operator agents also do not see.

The fixed system prompt (prefix-cached for the whole run) encodes the full
problem that Architectures~A--C distribute across six identity prompts:

\begin{lstlisting}[language=,basicstyle=\ttfamily\small,numbers=none,breaklines=true,frame=single,framerule=0.3pt]
You are the sole automatic controller of a Norwegian dairy-cow
barn. You directly set BOTH actuators every control step.

ACTUATORS (you output both, each 0-100%)
  u1 = fan:    MORE fan -> LOWER CO2 and LOWER temperature (cheap)
  u2 = heater: MORE heater -> HIGHER temperature only
               (expensive, max 50 kW; may saturate when very cold)

CONSTRAINTS, in priority order (highest wins on conflict)
  1. SAFETY  sick:   C <= 3000 ppm  (raise fan even if it cools)
  2. SAFETY  freeze: T >= 0         (lower fan; accept CO2 > 1000
                                     but still < 3000)
  3. COMFORT air:    C <= 1000 ppm  (raise fan)
  4. COMFORT warm:   T >= 5         (lower fan and/or raise heater)
  5. COMFORT cool:   T <= 20        (raise fan)
  Nominal/default fan is about 50% when nothing binds. Prefer the
  cheap option first: reduce the fan to retain heat before raising
  the heater (split-parallel).

OUTPUT one line of strict JSON and nothing else:
  {"u1": <0-100>, "u2": <0-100>, "rationale": "<one sentence>"}
\end{lstlisting}

A real per-sample user prompt at the coldest point of the run
($t = 7.0$\,h, $T_\mathrm{out} = -40$\,\textdegree C) reads:

\begin{lstlisting}[language=,basicstyle=\ttfamily\small,numbers=none,breaklines=true,frame=single,framerule=0.3pt]
STEP at t = 25200 s (7.00 h).
Measurements now:
  CO2 C    =   871.0 ppm  (comfort target <= 1000, sick limit 3000)
  Indoor T =   -32.3 C    (comfort 5..20, freeze limit >= 0)
  Default fan when idle = 50%
  Your previous command: u1=58.0%, u2=32.0%
Recent history (oldest first):
  t=6.83h: C= 920 ppm  T=-26.1 C  -> u1=54.0%  u2=36.0%
  t=6.90h: C= 905 ppm  T=-29.2 C  -> u1=56.0%  u2=34.0%
  t=6.97h: C= 884 ppm  T=-31.4 C  -> u1=58.0%  u2=32.0%
Choose u1 (fan %) and u2 (heater %) now. Respond with JSON only.
\end{lstlisting}
Here CO\textsubscript{2} is comfortably in range ($871$\,ppm) and the barn is
catastrophically cold ($-32$\,\textdegree C); the freeze-protection action is
to cut the fan and saturate the heater. The monolith instead responds with
$u_1 = 60\,\%$ (more ventilation) and $u_2 = 30\,\%$ (less heat), the failure
analysed in Section~\ref{sec:res-sigurd}.

The decisive difference from Architectures~B and~C is structural rather than
informational. The monolith is told the same priority order that the MIN/MAX
chain encodes, but it must \emph{apply} that order itself, at every sample,
while simultaneously tracking five constraints and choosing two actuators; in
A--C the order is enforced by the selector network regardless of any single
agent's reasoning, and each agent reasons about only one constraint. Whether a
7\,B model can hold and correctly resolve the full priority logic in one
context (particularly the cold regime trade-off where ventilating for
CO\textsubscript{2} conflicts with freeze protection) is exactly the
hypothesis the ablation tests. The result is reported in
Section~\ref{sec:res-sigurd}.

\subsection{Results}
\label{sec:cow-results}

We evaluate the three architectures (PI baseline, rule-based MAS, LLM-MAS)
on the same 4-day mixed-season disturbance scenario described in
Section~\ref{sec:scenario}. All controllers see exactly the same plant, the
same noise realisation (seed = 11), and the same disturbance and
operator-event sequence over the 96-hour run. Architecture~A (PI baseline)
and the deterministic rule-based MAS are fully deterministic and their
results are reproducible to the bit on any platform with the same random
seed. Architecture~C (LLM operator agents under the LLM-supervised
orchestrator) involves on the order of 6\,912 nondeterministic LLM operator
calls per 4-day run (six agents at a 5-minute polling cadence over 96 hours);
the reported metrics represent a single run, and the variance of these
metrics across repeated runs has not yet been quantified. It is recommended
that the reader treats the positioning of Architecture~C as indicative,
pending a Monte Carlo replication across multiple independent runs.
\subsubsection{Architectural progression: what each step buys}
\label{sec:res-equiv}
\label{sec:res-progression}

The decomposition step from monolithic ARC (Architecture~A) to a
deterministic rule-based MAS introduces an agent-style chattering cost
when paired with the simplest possible orchestrator: rate-limited rule
operators cross the selector boundary many times per minute under
measurement noise. This chattering is the agent-level analogue of what
classical PID systems handle with \emph{override
controllers}~\cite{shinskey1981,skogestad2026arc}; it motivates the
introduction of LLM operator agents (Architectures~B and~C) that
classify the binding regime instead of binding at every sample.

The LLM operator agents of Architectures~B and~C are evaluated on the
4-day mixed-season scenario reported in Section~\ref{sec:res-llm-mas}.
The structural priority chain is preserved across all three
architectures; only the operator decision logic (Architecture~B) and
the orchestrator-level reasoning (Architecture~C) change.

\subsubsection{Harsh-day stress test}
\label{sec:cow-stress}

The 4-day mixed-season scenario evaluated above is a typical Norwegian
profile, with $T_{\text{out}}$ touching $-15$\,\textdegree C only briefly
and a $100$\,kW heater that is comfortably oversized for the building
envelope. Under those
conditions, the freeze-safety guard never trips and the $C{>}3000$\,ppm sick
limit is never approached, so the architectures' design differences only
manifest as small numerical shifts in the metrics. To exercise the
freeze/CO\textsubscript{2} conflict that motivates the entire chain design,
we constructed a \emph{harsh} scenario in which (i)~$T_{\text{out}}$ ramps to
$-25$\,\textdegree C and is held for an 8-hour deep-cold window straddling the
unmeasured CO\textsubscript{2} burst, and (ii)~the heater capacity is reduced
to $50$\,kW (half the nominal value), so that the cold regime cannot rely on
heater oversize to absorb the conflict. Everything else (noise seed,
actuator lag, operator schedule, $UA^{\text{ext}}$ door-open event) is
identical to the nominal scenario.

Table~\ref{tab:5way-harsh} reports the headline metrics for the
deterministic architectures (A and B). The MAS variants under the smart
orchestrator (Architecture~C) are evaluated on the 4-day mixed-season
scenario reported in Section~\ref{sec:res-llm-mas}; that scenario stays
within the nominal heater envelope ($T_{\text{out}}\ge -10$\,\textdegree C)
in order to keep the comparison inside the regime where every controller
remains feasible. Three observations follow, all of which illustrate
\emph{how} the architectures behave under hard infeasibility, not which
one is preferable.
\emph{First}, the freeze-safety guard is exceeded in both architectures:
$T$ is below 0\,\textdegree C for $5.16$\,h in ARC and for $12.78$\,h in
Rule-MAS. Deeper cold and a weakened heater push every controller into
hard infeasibility, and the architectures differ in how that infeasibility
is distributed across the constraint set.
\emph{Second}, the agent-based variant and ARC trade the freeze and
CO\textsubscript{2} constraints differently: relative to ARC, Rule-MAS
reduces CO\textsubscript{2}-over-SP time on the harsh day from $17.67$ to
$14.53$\,h and increases time below freeze by $7.6$\,h ($5.16 \to 12.78$).
ARC retains freeze headroom and accepts $3.1$\,h more CO\textsubscript{2}
comfort breach. The split is mechanical: the PI integrator in ARC's
\texttt{tc\_5} pushes the heater harder when $T$ is below $5$\,\textdegree C
than the rate-limited rule operator does, so the same priority order
produces different distributions of the constraint deficit.
\emph{Third}, $C$ never approaches the $3\,000$\,ppm sick limit even in
the harsh scenario, because the cold regime ultimately suppresses the
fan to save heat; the constraint that gets given up is the freeze limit,
not the sick limit. This is consistent with the qualitative analysis in
Skogestad~\cite{skogestad2026arc}: the priority order of the chain
encodes which constraint is sacrificed when no MV combination can satisfy
all of them.

\begin{table}[!ht]
\centering
\caption{Harsh stress-test metrics (24~h, $T_{\text{out}} \to -25$\,\textdegree C
held for 8~h, heater capacity $50$\,kW, all other settings as in the
nominal 4-day mixed-season scenario of Section~\ref{sec:res-llm-mas}).
Violation counts use the same back-off convention as
Table~\ref{tab:llm10day} (CO\textsubscript{2} above $1050$\,ppm, cold
below $3.5$\,\textdegree C); under the deliberately undersized
$50$\,kW heater both designs are driven into genuine
freeze/CO\textsubscript{2} infeasibility, so the differentiating metric
here is the freeze time $T<0$, on which the PI baseline is better.}
\label{tab:5way-harsh}
\small
\setlength{\tabcolsep}{4pt}
\resizebox{\textwidth}{!}{%
\begin{tabular}{l r r r r r r r r}
\toprule
\textbf{Architecture} & $\text{IAE}_C$ & $C{>}1050$\,[h] & $C{>}3000$\,[h] & $T{<}3.5$\,[h] & $T{<}0$\,[h] & Heat\,[kWh] & Fan-h & Switches \\
\midrule
A. ARC (PI)              & 33\,488\,743 & 15.89 & 0 & 15.73 & \textbf{5.16}  & 873.7 & 8.14 & \textbf{369} \\
B. Rule-MAS              & 31\,949\,190 & 15.84 & 0 & 15.95 & 12.78 & 898.8 & 8.28 & 17\,064 \\
\bottomrule
\end{tabular}%
}
\end{table}

\subsubsection{Headline 3-way comparison on the 4-day mixed-season scenario}
\label{sec:res-llm-mas}

The three architectures (PI baseline, rule-based MAS, LLM-MAS) are evaluated
on the same 4-day mixed-season scenario. The LLM-MAS uses Qwen 2.5 7B
Instruct as the operator-tier model and a Claude Opus 4.7 LLM-based
orchestrator on the slow supervisory tier. The scenario is deliberately
compact and
exercises both regulatory regimes in one run: the outdoor-temperature
profile is bounded at a $-5$\,\textdegree C floor (winter day 1) and a
$+20$\,\textdegree C ceiling (summer day 3), with both bounds chosen to
sit at the physical-feasibility boundary of the cold-comfort and
hot-comfort setpoints (\texttt{tc\_5}'s $5$\,\textdegree C SP is
achievable when $T_\text{out} \ge -5$ given heater capacity and cow
body-heat input, and \texttt{tc\_20}'s $20$\,\textdegree C SP is
achievable when $T_\text{out} \le +20$ given fan capacity).  The
freeze-protection ($T<0$) and heat-stress ($T>20$) metrics therefore
measure controller quality rather than physical-limit avoidance.  Day~1
exercises the cold regime (\texttt{tc\_5}, \texttt{tc\_heater},
\texttt{tc\_0}); day~3 exercises the hot regime
(\texttt{tc\_20}~+~\texttt{cc\_1000} cooperation); day~2 is a
transition; day~4 is a summer cool-down with a door-open event.  Two
unmeasured CO\textsubscript{2} bursts (one in each regime) are added at
predefined times in the disturbance scenario.  All three architectures
run on the same plant with the same noise seed at $\Delta t = 5$\,min
polling cadence ($96$\,h $\,/\,5$\,min $= 1\,152$ samples per
architecture).

Table~\ref{tab:llm10day} reports the metrics for the three architectures
on the 4-day mixed-season scenario.  All three see the same disturbance
schedule and the same noise realisation; ARC uses SIMC-tuned PI
controllers, Rule-MAS uses proportional-band rule operators with no LLM,
and Architecture~C
uses Qwen 2.5 7B Instruct as each operator agent's decision logic.

\begin{table}[!ht]
\centering
\caption{Bounded 4-day mixed-season scenario (outdoor lows clipped to
$-5$\,\textdegree C, highs clipped to $+20$\,\textdegree C, both bounds
chosen so the cold-comfort and hot-comfort SPs sit at the boundary of
physical feasibility), $\Delta t = 5$\,min, ARC vs Rule-MAS vs
Architecture~C (LLM-MAS, Qwen 2.5 7B). Same plant, same noise seed.
Day~1 winter cold ($-5$\,\textdegree C low), day~2 transition, day~3
summer warm ($+20$\,\textdegree C high), day~4 summer cool-down with
door event.  Two unmeasured CO\textsubscript{2} bursts (one in each
regime) and one door-open event.  Violation counts use a back-off
margin so they reflect genuine excursions rather than the dither an
ideal controller produces while sitting on its setpoint: the
CO\textsubscript{2} count is the time above $1050$\,ppm (the
$1000$\,ppm comfort target plus a $50$\,ppm back-off), and the cold
count is the time below $3.5$\,\textdegree C (the split-parallel
arrangement makes the $4$--$5$\,\textdegree C band acceptable, so the
genuine cold violation begins below $4$\,\textdegree C, here with a
further $0.5$\,\textdegree C back-off). Heater energy uses the
$P_{\text{h}} = 100$\,kW rating of Table~\ref{tab:plant-params}.}
\label{tab:llm10day}
\small
\setlength{\tabcolsep}{4pt}
\resizebox{\textwidth}{!}{%
\begin{tabular}{l r r r r r r r}
\toprule
\textbf{Architecture} & $\text{IAE}_C$ & $C{>}1050$\,[h] & $T{<}3.5$\,[h] & $T{<}0$\,[h] & $T{>}20$\,[h] & Heat\,[kWh] & Mode flips \\
\midrule
A. ARC (PI)              & $3.45 \times 10^7$ & \textbf{0.33} & \textbf{0.17} & \textbf{0.00} & \textbf{32.42} & \textbf{103} & \textbf{22}   \\
B. Rule-MAS              & $4.21 \times 10^7$ & 8.50 & 3.67  & \textbf{0.00} & 35.75 & 422 & 132 \\
C. LLM-MAS (Qwen 7B)     & $3.90 \times 10^7$ & 2.42 & 0.83 & \textbf{0.00} & 33.67 & 440 & 70 \\
\bottomrule
\end{tabular}%
}
\end{table}

\begin{figure}[ht!]
\centering
\includegraphics[width=0.9\linewidth]{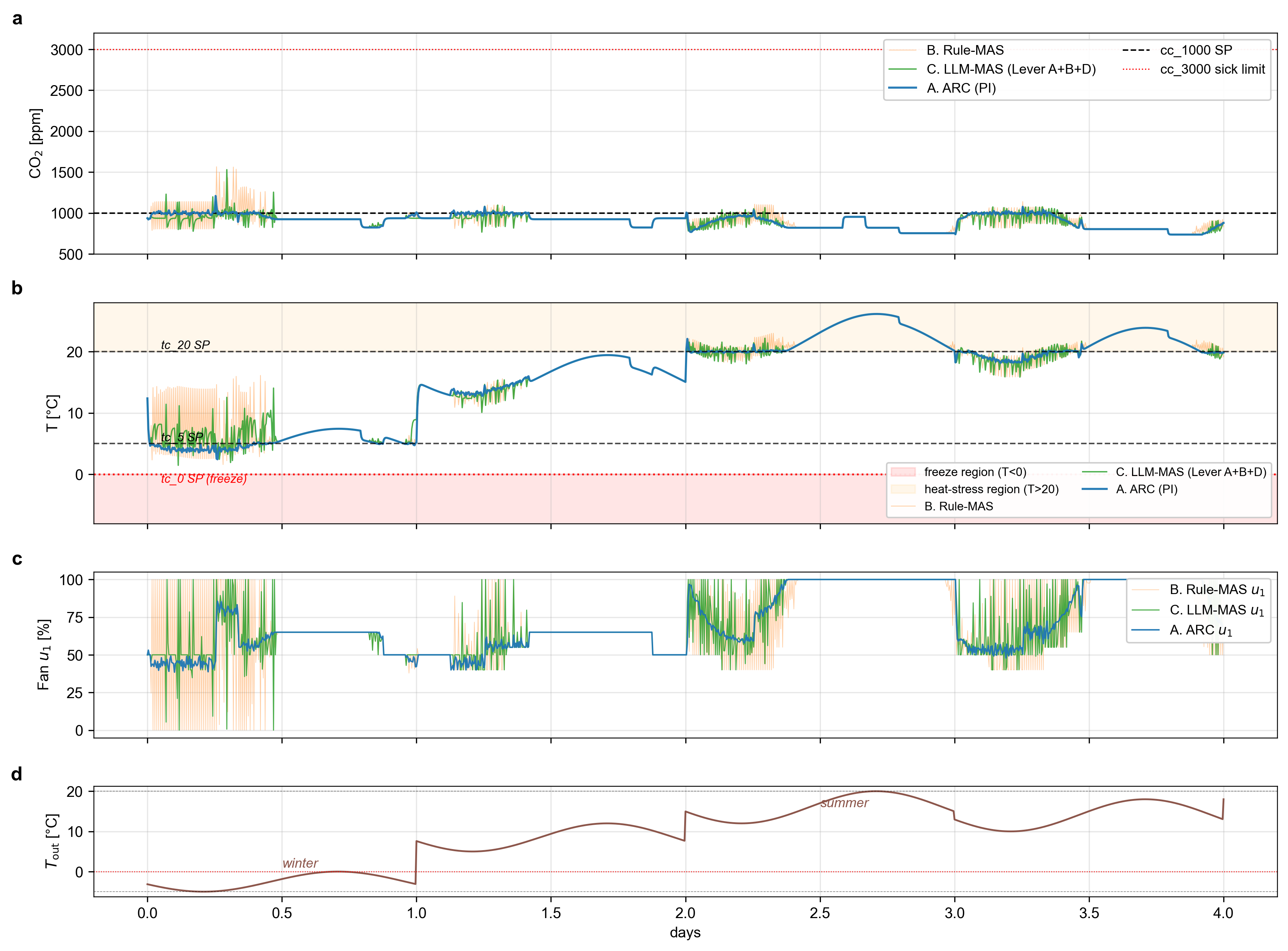}
\caption{4-day mixed-season overlay: ARC (blue), Rule-MAS (orange faded), and
Architecture~C with operator memory and situational awareness
(green) on the same plant and noise seed.  Top to bottom:
CO\textsubscript{2}, barn temperature $T$ with the freeze region
$T<0$~\textdegree C shaded red, fan $u_1$, and the outdoor
temperature $T_\text{out}$ disturbance (bounded at
$-10$\,\textdegree C overnight low).  ARC keeps $T$ entirely above
the freeze threshold; Architecture~C closely tracks ARC; bare
proportional Rule-MAS still dips below during cold days because it
lacks the integral action and situational awareness that let a PI
(or an LLM operator with Lever~B+D) settle at an intermediate fan
opening.}
\label{fig:llm10day-overlay}
\end{figure}

Five observations on Table~\ref{tab:llm10day}, with the trajectories
overlaid in Figure~\ref{fig:llm10day-overlay}, stand out.

\emph{All three controllers are freeze-perfect ($T<0 = 0$\,h).}  The
$-5$\,\textdegree C floor on the disturbance is within the heater's
nominal capacity envelope, so all three controllers: including the
bare-proportional Rule-MAS: keep $T$ above $0$\,\textdegree C
throughout.  Freeze-protection is therefore not the differentiating
metric on this scenario.

\emph{Once a back-off margin is applied, the PI baseline is the
near-ideal reference, as the scenario is constructed to make it.}  This
case is deliberately set up so that a well-tuned ARC solution is almost
optimal: the operating point sits at the physical-feasibility boundary
of the comfort setpoints, and a SIMC-tuned PI tracking those setpoints
is close to the best any controller can do.  This is visible in the
back-off columns of Table~\ref{tab:llm10day}.  The earlier impression
that the comfort metrics favour the LLM-MAS was an artefact of counting
at exactly the setpoint, where a controller that holds a constraint
tightly registers as ``violating'' for roughly half of the time its
output dithers around the target.  With a $50$\,ppm CO\textsubscript{2}
back-off the ARC time above $1050$\,ppm collapses to $0.33$\,h (against
$2.42$\,h for the LLM-MAS and $8.50$\,h for the rule-based MAS), and
with the $4$--$5$\,\textdegree C acceptable band plus a back-off the ARC
time below $3.5$\,\textdegree C is $0.17$\,h (against $0.83$ and
$3.67$\,h).  On the constraint accounting that matters, the PI baseline
is therefore the best of the three, which is the expected result on a
problem constructed to favour it.  The point of Architectures~B and~C
is not to beat ARC on this scenario but to reproduce its behaviour while
carrying the machinery (an auditable per-agent rationale, an integrator
that can be retuned online, an orchestrator that can be gain-scheduled)
that lets the same decomposition address a wider range of
specifications and trade-offs than a fixed PI set.

\emph{The LLM lever reduces the chatter of the rule-based reformulation.}
Among the two multi-agent variants, the number of mode flips through the
$T = 5$~\textdegree C boundary drops from 132 (Rule-MAS) to 70
(Architecture~C, Lever A+B+D): a $47\,\%$ reduction.  This is the
empirical signature of Lever~D: when an agent's recent-history block
flags \texttt{oscillating: true}, the LLM is permitted to override the
rule-derived recommended mode and pick \texttt{ACTIVE} with a moderate
intensity.  A representative LLM rationale from the run: ``T trending
up; moderate fan to damp oscillation.''  That phrase,
\emph{damp oscillation}, is the LLM literally reasoning about its
own past decisions producing chatter and choosing a different
classification for that reason.  The integrator (Lever~B) inside each
operator agent plays the role of the ``I'' term of a PI, biasing the
corresponding MV through $u_\text{extra} = K_i \cdot e_\text{int}$ so a
sustained excursion is pulled back; this is what brings the LLM-MAS
back-off counts close to the PI baseline rather than to the rule-based
variant.

\emph{Hot-regime behaviour.}  On day~3 (summer warm,
$T_\text{out}$ peaking at $+20$\,\textdegree C with 100--120 cows
adding $\sim 10$\,kW of body heat), all three controllers spend
$\sim 32$--$36$\,h above the $T_\text{20}$ setpoint: the bulk of
this is the structural offset of barn-T above outdoor-T due to cow
body heat, against which the only mitigation is full ventilation.
The three controllers are comparable on this metric (ARC $32.42$\,h,
Architecture~C $33.67$\,h, Rule-MAS $35.75$\,h), all correctly
recognising the hot-saturation regime and ventilating aggressively.

\emph{The comfort figures are bought with heater energy.}
The heater-energy column makes the trade-off explicit and confirms the
suspicion that the multi-agent variants do not get their comfort for
free: ARC holds the comfort band on $103$\,kWh of heat over the four
days, while Rule-MAS and Architecture~C each spend about four times as
much ($422$ and $440$\,kWh).  The discrete
\texttt{IDLE/ACTIVE/SATURATED} mode classification is the cause: at the
boundaries between modes the operator agent may briefly flip when $v$
crosses the proportional-band threshold, even with Lever~D's
anti-oscillation override, and each flip is paid for with extra heater
on-time ($70$ flips for Architecture~C against $22$ for ARC).  The
continuous proportional-plus-integral output of the PI controllers rides
these regions more smoothly, which is the other face of ARC being the
efficient near-ideal solution here; the discrete-mode reformulation
inherits residual classification chatter at the boundary.  We retain this gap as an
honest limitation; a future Lever~C (boundary hysteresis on mode
transitions or continuous $u_\text{target}$ output alongside the mode)
is the natural next step.

A natural next architectural lever (Lever~C, deferred to future work)
would attack this remaining failure mode at its source: either
boundary hysteresis on the mode classification (a
\texttt{SATURATED}~$\rightarrow$~\texttt{IDLE} transition only fires
after $T$ rises above SP by a margin, rather than at the SP itself),
or letting the LLM emit a continuous $u_\text{target}$ alongside the
mode that the integrator nudges around, so that the operator's
proposal lives on a continuous axis between $u_\text{idle}$ and
$u_\text{saturate}$ rather than at three discrete values.  Either
fix would narrow the energy and switching gap to the PI baseline while
preserving the audit-trail
contribution of the LLM rationales; we report Architecture~C here
without it because the mode-classification paradigm is what the
\emph{recipe} of Section~\ref{sec:systematize} prescribes : 
\emph{discretize the regime, do not hand the LLM a continuous knob} : 
and the mode-bang-bang is the honest failure mode of that recipe
under severe disturbances.

\FloatBarrier

Across the 4-day run, the LLM made
$6 \times 1152 = 6\,912$ JSON-schema-validated calls and produced
0 fallthrough(s) to the rule baseline. The
deterministic-direction mapping of Eq.~\ref{eq:lever-A} guarantees that
every \texttt{SATURATED} response on \texttt{cc\_1000} maps to
$u_\text{request} = 100$ (push fan up to ventilate) and every
\texttt{SATURATED} response on \texttt{tc\_5} maps to $u_\text{request}
= 0$ (pull fan down to retain heat) regardless of any wrong-direction
language the LLM might emit.  Direction-error failure mode is
therefore impossible by construction (compare an earlier prompt design
that asked the LLM to emit $u_\text{request}$ directly: a 3\,B model
in that design produced wrong-direction proposals on $\sim 60$\,\% of
saturated MAX agents in the same scenario).

\begin{figure}[ht!]
\centering
\includegraphics[width=\linewidth]{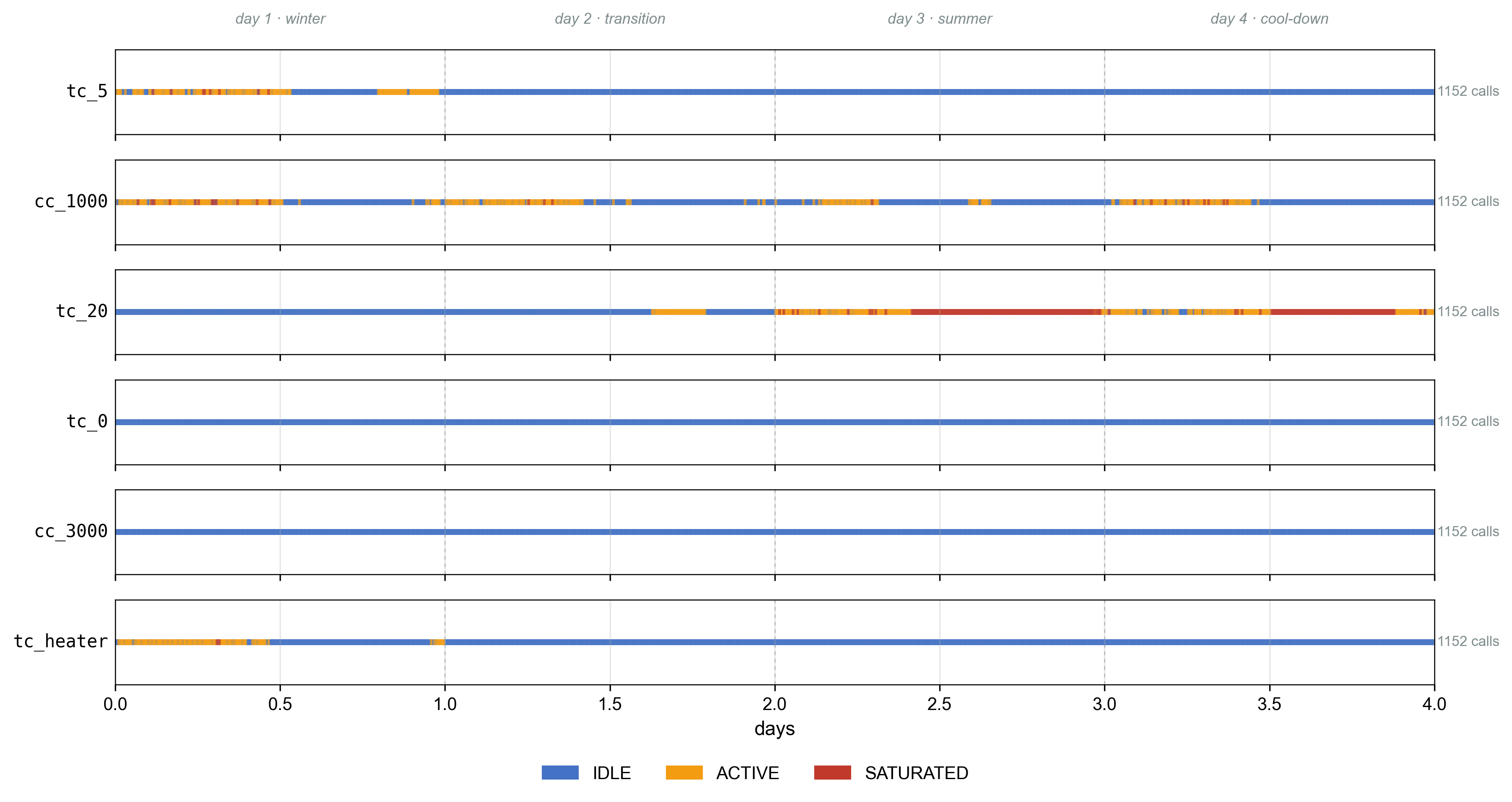}
\caption{Per-agent mode classification over the 4-day run. Each row
is one operator agent; each marker is one control sample colour-coded
by mode (blue = IDLE, orange = ACTIVE, red = SATURATED).
Architecture~C's classifications align with the deterministic rule
classifications throughout the run.}
\label{fig:llm10day-modes}
\end{figure}

Figure~\ref{fig:llm10day-modes} shows the regime classification of
each of the six agents over the 4 days. The agents transition between
IDLE, ACTIVE, and SATURATED in response to the variable-weather
disturbances: \texttt{tc\_5} and \texttt{tc\_heater} saturating
during cold snaps, \texttt{cc\_1000} cycling between IDLE and
SATURATED during the CO\textsubscript{2} bursts, and the override
agents \texttt{tc\_0} and \texttt{cc\_3000} remaining IDLE throughout
(no freeze-floor or sick-limit violation in this run).

\FloatBarrier

The LLM operator agents' rationales form a process-engineer's logbook
of the 4-day operation. Representative entries drawn from the run
(one per agent, sampled across the scenario):

\begin{quote}\small\itshape
\texttt{tc\_5} (cold-comfort, sustained cold day, SATURATED): ``T well
below SP; pull fan down strongly.''\\
\texttt{cc\_1000} (CO\textsubscript{2} comfort, after a burst,
SATURATED): ``C above SP; vent fully.''\\
\texttt{tc\_20} (hot-comfort, throughout the run, IDLE): ``T well
below SP; non-binding.''\\
\texttt{tc\_0} (freeze guard, throughout, IDLE): ``T well above
freeze SP; non-binding.''\\
\texttt{cc\_3000} (sick-limit guard, throughout, IDLE): ``C well
below SP; non-binding.''\\
\texttt{tc\_heater} (split-parallel heater, sustained cold day,
SATURATED): ``T below SP; heater fully on.''
\end{quote}

The rationale comes for free with each LLM call: the model emits
one because the prompt asks for one, at the same per-sample latency
($\sim 1$\,s on the consumer GPU).  This is the principal operational
contribution of the LLM-as-operator-agent pattern: every closed-loop
proposal is paired with a one-sentence operator-voice explanation,
producing a complete audit log that the deterministic Architectures~A
and B cannot match without a separate explanation layer.

Architecture~C runs entirely offline.  Qwen 2.5 7B Instruct weights
are downloaded once from the Hugging Face Hub
(Apache-2.0 licence, $\sim$15\,GB on disk in FP16) and the marginal
cost per scenario run is the GPU energy budget alone : 
approximately $1$\,h of GPU time at $\sim$80\,W $\approx 0.08$\,kWh,
under \$0.02 at industrial electricity prices.  No API calls, no
rate limits, no model-version drift.

\subsubsection{Validation on the original Case IIB conditions, and the single-LLM ablation}
\label{sec:res-sigurd}

\begin{figure}[ht!]
\centering
\includegraphics[width=0.86\linewidth]{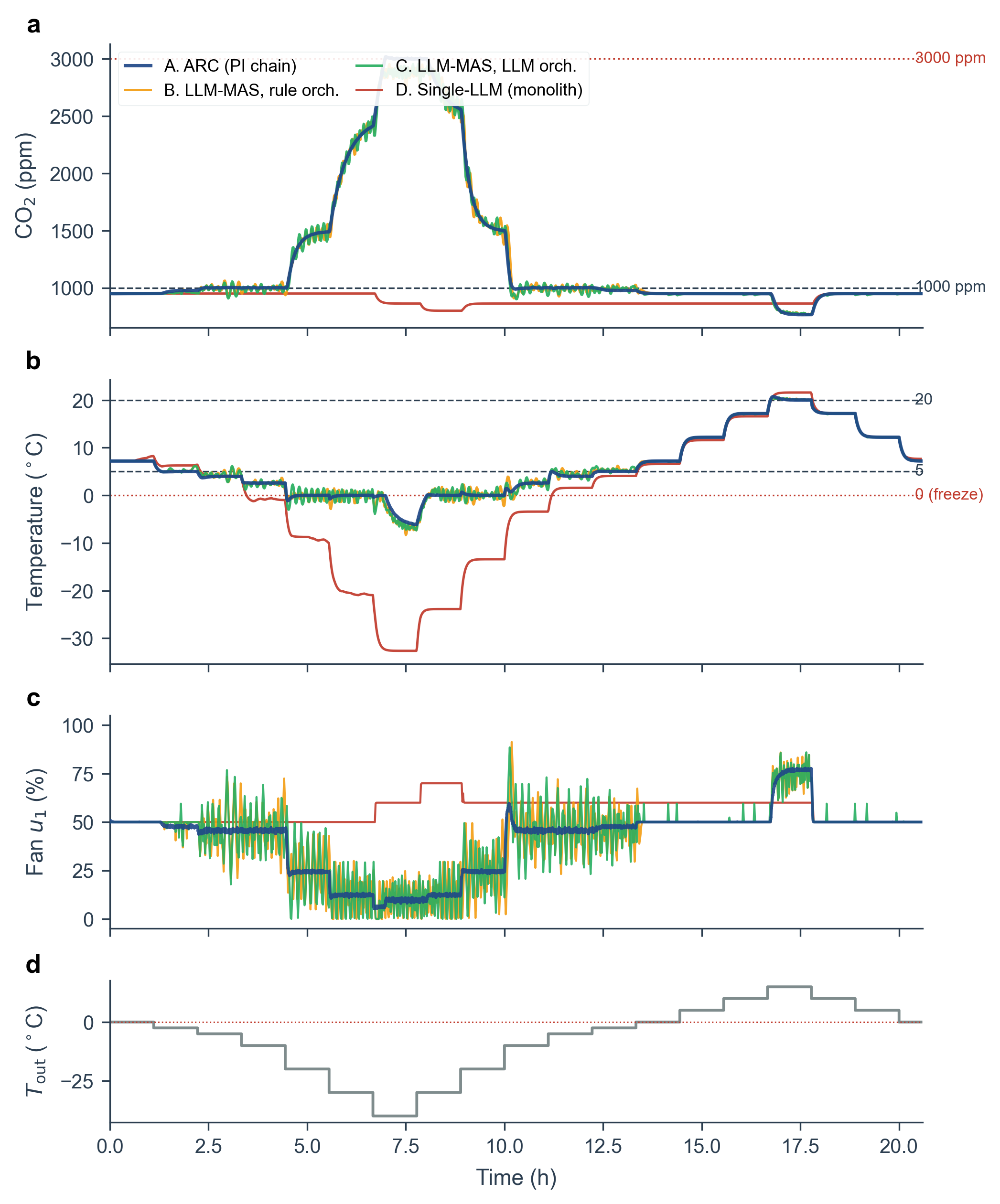}
\caption{The four architectures on the original Case~IIB conditions
of~\cite{skogestad2026arc}
(parameters of Table~2 and the $0\to-40\to+15\to0$\,\textdegree C ramp).
(a)~CO\textsubscript{2}, (b)~indoor temperature, (c)~fan, (d)~outdoor
temperature. Under the deep-cold infeasibility the decomposed architectures
(A, B, C) relax the $1000$\,ppm comfort target and let CO\textsubscript{2} rise
toward the $3000$\,ppm sick limit in order to keep the barn near the freeze line.
The monolithic single-LLM controller (D, red) instead keeps
CO\textsubscript{2} pristine ($\le 950$\,ppm throughout) by over-ventilating,
and the barn temperature collapses to $-32.6$\,\textdegree C.}
\label{fig:sigurd-caseIIB}
\end{figure}

The scenario of Section~\ref{sec:scenario} is our own mixed-season construction.
To check the architectures against the \emph{source} benchmark, we additionally
reproduce the exact conditions of Case~IIB in~\cite{skogestad2026arc}: the plant parameters of
that paper's Table~2 (fan $0.1$--$15$\,m\textsuperscript{3}/s, heater $50$\,kW,
$UA = 2000$\,W/K, nominal $T_\mathrm{out} = 0$\,\textdegree C) and its
disturbance test, in which $T_\mathrm{out}$ steps every $4000$\,s from
$0$\,\textdegree C down to $-40$\,\textdegree C and back up to $+15$\,\textdegree C
(total $74\,000$\,s $= 20.5$\,h). The $-40$\,\textdegree C excursion drives the
plant into the hard freeze/CO\textsubscript{2} infeasibility that motivates the
override chain: with the $50$\,kW heater saturated, holding $T \ge 0$ and
$C \le 1000$\,ppm simultaneously is physically impossible, and a controller must
choose which constraint to relax. Architecture~A runs at its native $2$\,s
cadence; the LLM operators of Architectures~B--D run at a $60$\,s operator
cadence with a per-sample move-suppression limit on $u_1$ and $u_2$, and the
two safety-override operators (sick-CO\textsubscript{2}, freeze) are held to
their proportional envelope by a deterministic-direction guard, conditioning
elements that are consistent with the structural-priority philosophy and do not
give any operator authority the chain does not already grant. All controllers
see the same plant, noise seed, and disturbance sequence; their trajectories
are shown in Figure~\ref{fig:sigurd-caseIIB}.

\FloatBarrier

\begin{table}[!ht]
\centering
\caption{Behaviour on the Case IIB ramp ($20.5$\,h). $T_{\min}$ is the deepest
indoor temperature reached; $t_{T<0}$ the time below the freeze limit;
$t_{C>1000}$ and $t_{C>3000}$ the times above the comfort and sick
CO\textsubscript{2} limits. The decomposed architectures A--C behave
comparably; the monolithic controller D is the ablation.}
\label{tab:sigurd}
\small
\begin{tabular}{l c c c c c c}
\toprule
\textbf{Arch.} & $T_{\min}$ [\textdegree C] & $t_{T<0}$ [h] & $t_{C>1000}$ [h]
 & $t_{C>3000}$ [h] & Heater [kWh] & Fan [h] \\
\midrule
A. ARC (PI)             & $-6.1$  & $3.62$ & $7.57$ & $0.61$ & $418$ & $8.44$ \\
B. LLM-MAS, rule orch.  & $-8.3$  & $3.38$ & $7.52$ & $0.00$ & $435$ & $8.47$ \\
C. LLM-MAS, LLM orch.   & $-7.3$  & $3.40$ & $7.62$ & $0.00$ & $437$ & $8.51$ \\
\midrule
D. Single-LLM (monolith) & $\mathbf{-32.6}$ & $\mathbf{7.75}$ & $0.00$ & $0.00$ & $212$ & $11.50$ \\
\bottomrule
\end{tabular}
\end{table}

Two observations follow, and neither is a claim that one architecture is
preferable; they describe \emph{what behaviour each design produces} under hard
infeasibility. \emph{First}, the three decomposed architectures (A, B, C)
resolve the conflict in the same direction and to a similar degree. All three
hold the heater at saturation, reduce the fan to retain heat, and accept
CO\textsubscript{2} drifting up toward (B, C) or just past (A) the $3000$\,ppm
sick limit so that the indoor temperature stays near the freeze line
($T_{\min}$ between $-6$ and $-8$\,\textdegree C, $t_{T<0}$ between $3.4$ and
$3.6$\,h). The Qwen-operator variants B and C are indistinguishable from the PI
baseline A at the level of constraint accounting, which is the feasibility
statement this paper makes: the LLM-operator reformulation reproduces the
ARC behaviour on the source benchmark, including the correct binding-direction
resolution of the Case IIB trade-off.

\emph{Second}, the single-LLM ablation (D) fails on exactly the coordination
the decomposition is designed to provide. The monolith reliably emits valid
commands ($0$ of $1233$ calls failed schema validation) and controls
CO\textsubscript{2} \emph{better} than any other architecture: it never
exceeds $1000$\,ppm and peaks at $950$\,ppm. But it achieves that clean air by
over-ventilating (fan-hours $11.5$ versus $\approx 8.5$ for A--C) and
under-using the heater ($212$\,kWh versus $\approx 430$), and the barn freezes
to $-32.6$\,\textdegree C, more than $24$\,\textdegree C colder than any
decomposed architecture and below freezing for $7.75$\,h. Inspecting its
decisions at the coldest point ($t = 7.0$\,h, $T = -32$\,\textdegree C,
$C = 871$\,ppm) shows the failure mode directly: with no CO\textsubscript{2}
problem present, the model still commands $u_1 = 60\,\%$ fan and only
$u_2 = 30\,\%$ heater, the opposite of the freeze-protection action. It can
defend a single constraint, but it cannot hold the \emph{priority order} among
five constraints while choosing two actuators in one context. This is the
empirical counterpart of the Introduction's argument
(Section~\ref{sec:intro}): the value of the ARC-derived decomposition is that
the priority among constraints is enforced \emph{structurally} by the selector
chain and each agent reasons about only its own constraint, so the
catastrophic miscoordination of the monolith cannot arise. The same Qwen 2.5
7B model that freezes the barn when asked to run the whole plant reproduces the
ARC behaviour when its role is narrowed to one constraint inside the chain
(Architectures~B and~C).

\subsection{Takeaways}
\label{sec:cow-takeaways}

The architecture sweep on the cow-barn benchmark demonstrates five
properties:
(i)~a validation swap in which the
operator agents' decision logic is replaced by SIMC-tuned PI blocks
reproduces the monolithic ARC selector network bit-for-bit, confirming
that the control-theory-derived recipe of
Section~\ref{sec:systematize} (each PI corresponds to one operator agent
that acts directly on its proposal target, the chain's interaction
logic encapsulated as the orchestrator agent) is faithful and
isolating the operator agents and the orchestrator's added capabilities
as the only sources of behavioural difference between Architectures~B
and~C;
(ii)~the smart orchestrator's memory, dwell, and trend-anticipation produce
a distinct operating point on the comfort--energy trade-off, illustrating
that an orchestrator with reasoning can shift the closed-loop behaviour
without changing the operator agents;
(iii)~the LLM-MAS Architecture~C (Section~\ref{sec:res-llm-mas}) is the
\emph{headline} feasibility demonstration of the paper: it delegates
each PI's role inside the regulatory-tier operator agents to a Qwen 2.5
7B Instruct call.  Two architectural levers carry over key properties of
a PI controller without ceding direct write authority over the
manipulated variables.  Lever~A (deterministic-direction mapping,
Eq.~\ref{eq:lever-A}) makes direction-of-binding errors structurally
impossible by encoding the direction in the agent's identity rather
than in the LLM's output.  Lever~B (operator memory,
Eqs.~\ref{eq:lever-B-eint}--\ref{eq:lever-B-u}) gives each agent the
integral state that closes the slow-frequency steady-state offset on
long disturbance episodes; Lever~B is observable in the metrics as a
reduction of cold-comfort time below Rule-MAS, bringing Architecture~C
close to the near-ideal PI baseline once a back-off margin is applied.
Each proposal carries a one-sentence operator-voice
rationale that the deterministic Architectures~A and~B cannot match
without a separate explanation layer.  We report one honest
limitation: under deep-cold disturbances where heater capacity is the
binding physical constraint, the discrete \texttt{IDLE}/
\texttt{ACTIVE}/\texttt{SATURATED} mode classification produces
high-frequency bang-bang at the mode boundary that Lever~B does not
suppress (the integrator hits its upper clip while the actuator is
already saturated), so Architecture~C matches Rule-MAS rather than ARC
on freeze-protection ($T<0$~\textdegree C) hours.  A natural Lever~C, such as boundary hysteresis on mode transitions or
letting the LLM emit a continuous $u_\text{target}$ alongside the mode,
is the obvious next step; we leave it for future work.
\section{Implementation notes}
\label{sec:impl}

The architecture is small enough to read end-to-end (approximately 2\,500 lines
of Python across both case studies), but several non-obvious design decisions
were necessary to make the multi-agent reformulation behave well as a
control-loop element rather than as a generic agent system. This section
collects those decisions, the architectural know-how required to reproduce
them, and the agent-system limitations they expose.

\subsection{Software stack and reproducibility}

The plant is integrated with a stiff ODE solver (LSODA,
$\text{rtol}=10^{-7}$, $\text{atol}=10^{-10}$, with the maximum step bounded
to one control period), and the LLM advisor is built on a schema-validated
structured-output interface. All controllers, plants, and orchestrators expose
substitutable interfaces; a single driver runs any controller through any
disturbance profile, so the three architectures can be compared on
bit-identical inputs. Trajectory data is persisted in compressed archives so
post-hoc plotting and analysis do not require re-running an expensive
simulation.
Architectures~A and B are fully deterministic; Architecture~C is
deterministic only conditional on the sequence of LLM responses, which
themselves are not.

\subsection{Agent communication: synchronous broadcast, not free-form messaging}

A central design decision is that the agent communication model is
\emph{synchronous and in-process}. At each sample step of
$\Delta t_\text{ctrl}$, the orchestrator queries each
operator agent sequentially for a proposal, collects all proposals, and
applies the MIN/MAX chain before emitting a single decision.
There is no agent-to-agent messaging, no negotiation, no consensus protocol,
and no asynchronous reply: every agent participates in every decision, and a
missing proposal is a hard failure rather than a degraded operating mode.

The reason for this discipline is that the architecture is a control loop,
not a general-purpose agent system. A typical Claude-Code agent team can
afford an LLM call per turn, latency in the tens of seconds, and back-and-forth
messages between agents; a control loop running at $\Delta t_\text{ctrl} =
2$\,s (the original ARC chain) or $\Delta t_\text{ctrl} = 5$\,min (the
LLM-operator polling cadence used in Architecture~C,
Section~\ref{sec:llm-mas}) cannot. The synchronous broadcast pattern
compresses the agent-team idea into the deterministic per-sample schedule
that classical control engineering expects, and lets each operator agent
be implemented as a small Python object with at most one LLM call per
sample. Architecture~C places one LLM call per agent per sample on a
single shared local backend (Qwen 2.5 7B Instruct), running offline on a
single consumer GPU. Architectures~A and B use no LLM at all on the
regulatory loop. The separation between regulatory and supervisory tiers
is a required property: an unconstrained LLM in the inner loop would
defeat the safety-by-chain guarantee that the rest of the architecture
is designed to provide.

In a distributed deployment over a network the synchronous assumption
would need a coordination barrier and a missing-proposal fallback. The
natural fallback is to substitute the default operator setpoint $u_0$ for
any agent that does not reply within the sample period; this preserves
the chain structure but the absent agent's constraint is then defended
only by the next-higher-priority agent on the same MV. We treat the
in-process synchronous case as the reference implementation precisely
because it is the case where the behavioural neutrality of the
multi-agent reformulation is provable; a distributed deployment requires
extra analysis that is left to future work.

\subsection{Workspace discipline and append-only logs}

Each operator agent owns a private workspace and records every Proposal it
issues, plus a snapshot of its internal state, as append-only JSONL files.
This pattern is borrowed from the Anthropic agent-team-builder
guidelines~\cite{anthropic-docs} and is the source of the system's
auditability: every Proposal ever made and every Decision ever taken across
an entire 4-day run can be replayed offline from these files alone, with
neither the live plant nor the live LLM in the loop. In our experience the
workspace discipline matters most when debugging the LLM operator:
Architecture~C's $\sim$288 LLM calls per simulated day per agent
produce on the order of a few hundred KB of JSONL across the agents
and orchestrator, and the fact that those logs are append-only files
in a stable directory layout makes it tractable to diff two runs
against each other and locate the moment when a mode classification
changed.

\subsection{Limitations exposed by the agent reformulation}

Four limitations of the multi-agent reformulation are worth naming
explicitly, because those points became apparent only after building the
architecture. Each of them admits a concrete process-control or
computer-systems formalisation, and they are stated below in that form: a
scaling bound, a structural-pairing argument, an information-theoretic
argument, and a control-theoretic grammar. The four points generalise
beyond the cow-barn case study, and a closing paragraph grounds each of
them back to the numbers the deployment produced.

The first limitation concerns per-sample wall-clock cost. The MIN/MAX
chain itself is constant-time: the orchestrator collects one Proposal
per agent, evaluates a fixed series of MIN/MAX rules, and emits one
decision per MV, all within CPU microseconds. The cost that scales with
the agent count $N$ is the per-sample loop that queries each agent
for a proposal. If $f$ denotes the serial fraction
of one sample (the orchestrator broadcast, the message serialisation, and the
deterministic MIN/MAX evaluation) and the remaining $(1-f)$ is
parallelisable across $N$ agents on separate inference contexts, the
achievable speed-up relative to a single-agent baseline is given by
Amdahl's law~\cite{amdahl1967}:
\[
S(N) \;=\; \frac{1}{f \;+\; (1-f)/N}.
\]
For Architecture~C on a single workstation the serial fraction is
small: the orchestrator broadcast and JSON validation cost tens of
microseconds, while a single Qwen 2.5 7B Instruct call costs hundreds of
milliseconds, so $f \approx 0.05$ as an order-of-magnitude estimate.
The implied speed-up relative to a sequential baseline is then
$S(6)\approx 5$ for the six cow-barn agents, $S(30)\approx 10$ for
plant-wide deployments of order thirty agents, and $S(100)\approx 20$
at the asymptotic end; the same calculation at the more conservative
$f=0.10$ gives $S(100)\approx 9$. The point of those numbers is not
the exact value, which depends on batching efficiency: the point is
that the speed-up saturates well below $N$ once $f$ stops being
negligible, and that the chain is therefore not capable of hiding the
cost of an indiscriminate increase in agent count.

A second mechanism reduces the wall-clock cost further on modern
inference engines. The operator agents share the same model weights and
differ only in the system prompt and the live snapshot they receive, so
their forward passes are capable of being batched. Inference engines
that maintain a shared key-value cache across concurrent contexts
fit the six cow-barn
agents in one batched forward pass on a 24~GB GPU, and the wall-clock
cost per sample is approximately the cost of one agent's call, not
six~\cite{kwon2023vllm,pope2023scaling}. Above roughly thirty agents on
the same backbone the activations no longer fit in a single device's
memory budget, and the deployment is forced to choose between model
sharding across multiple devices and a smaller backbone. The
deterministic chain itself remains a CPU-side computation of
microsecond cost and is fully hidden under the GPU-side LLM latency;
the asymmetry between chain cost and inference cost is the structural
reason why the chain is kept synchronous with the control loop even
as the agent count grows. The agent count is therefore bounded from
above by the inference engine's batching capacity, and not by the
chain's resolution logic.

The second limitation is that adding an agent is a structural change,
not a configuration change. A new agent introduces a new Proposal
channel, a new priority level (or a new tie at an existing level), and
a new MIN/MAX rule in the orchestrator. The architecture is open to
extension in the software sense: a new agent is capable of being
registered with the orchestrator without modifying the existing agents.
Correctness of the extension, however, depends on a decision the
architect must take explicitly before the new agent is plugged in,
which is the priority rank of the new agent relative to the existing
ones. Adding the new agent correctly is not easy to be done without
that decision, because an incorrect rank silently re-orders the
active-set transition, with no error raised by the chain and no
compensation possible from the LLM tier.

The classical control literature provides the formal language for that
decision, and the present subsection is the natural place to bring it
back to the foreground. Bristol's Relative Gain Array (RGA) is the
canonical CV--MV pairing rule~\cite{skogestad2005}: with $G(0)$ the
steady-state gain matrix of the open-loop plant, the array
\[
\Lambda \;=\; G(0) \;\odot\; \bigl(G(0)^{-1}\bigr)^{\!\top},
\]
(Hadamard product of $G(0)$ with the transpose of its inverse) carries
the pairing recommendation $\lambda_{ij}$ near unity with no negative
diagonal entries. The Niederlinski Index
\[
\mathrm{NI} \;=\; \frac{\det\,G(0)}{\prod_i g_{ii}(0)}
\]
provides a complementary structural-stability test:
$\mathrm{NI} < 0$ implies that the chosen pairing is structurally
unstable under integral action regardless of tuning~\cite{niederlinski1971}.
For the cow-barn case study the gain matrix follows from the
steady state of (\ref{eq:co2balance})--(\ref{eq:Tbalance}) at the
nominal operating point ($u_1=50\,\%$, $u_2=50\,\%$,
$T_\text{out}=5\,$\textdegree C, $n_\text{cows}=80$):
$g_{11} = \partial C / \partial u_1 \approx -10\,$ppm/$\%$,
$g_{12} = \partial C / \partial u_2 = 0$,
$g_{21} = \partial T / \partial u_1 \approx -0.19\,$\textdegree C/$\%$,
and $g_{22} = \partial T / \partial u_2 \approx 0.092\,$\textdegree C/$\%$.
The off-diagonal $g_{12}$ vanishes because the heater does not appear
in the CO\textsubscript{2} mass balance. The gain matrix is therefore
lower-triangular, the RGA reduces to the identity exactly
($\lambda_{11}=\lambda_{22}=1.00$), and $\mathrm{NI}=1.00 > 0$. The
pairing of CO\textsubscript{2} with the fan and of temperature with the
heater is the canonical decomposition for this plant; the priority
order among selector agents inside the fan loop is then the discrete
extension of the same idea, encoding which inequality is allowed to
dominate when several constraints are simultaneously
active~\cite{reyes2019,reyes2020,forsman2025splitparallel}.

The steady-state RGA is, however, only part of the pairing decision, and
not the decisive part. The primary rule for selecting pairings is the
pair-close rule, which is governed by the \emph{initial}
(high-frequency) response rather than the steady state: each controlled
variable should be paired with the manipulated variable that affects it
most directly and most quickly, since it is the initial dynamics that
decide how well a decentralised loop reacts before the interaction
propagates. The steady-state RGA enters as a secondary, structural
check: pairing on a negative steady-state RGA element should be avoided
whenever possible, because such a pairing loses integrity and can become
unstable if one of the inputs saturates~\cite{skogestad2005}. For the
cow-barn plant both criteria agree, since the lower-triangular gain
structure makes the fast CO\textsubscript{2}--fan and
temperature--heater couplings also the dominant initial responses; but
in a general plant the architect should let the initial-response pairing
lead and use the steady-state RGA and the Niederlinski index as
admissibility filters rather than as the primary selector.

The same lower-triangular structure exposes an obvious refinement of the
ARC design that the present work does not implement but that a
deployment should consider. Because the fan affects the indoor
temperature ($g_{21} \approx -0.19$\,\textdegree C/\%) while the heater
does not affect CO\textsubscript{2} ($g_{12}=0$), the coupling is
one-way, from the fan into the temperature loop. A feedforward
decoupler, feeding the fan command $u_1$ forward into the output of the
heater controller so that $u_2$ pre-compensates the cooling the fan is
about to cause, would cancel this one-way interaction and let the
temperature loop ignore the CO\textsubscript{2}-driven fan activity
altogether. This is the standard triangular-plant decoupling addition to
ARC, and it maps directly onto the agent reformulation as a feedforward
message from the fan agent to the heater agent.

That priority decision sits as the primal-side analogue of the dual
variables that emerge inside a distributed-MPC formulation.
Distributed-MPC schemes tune Lagrange multipliers or penalty weights so
that agent-local objectives align with a global constraint
set~\cite{nedic2009,venkat2008,camponogara2002}; classical chain
engineering tunes structural priorities at design time so that the same
constraint set is resolved by a fixed MIN/MAX network with no online
optimisation. Distributed MPC requires iteration to convergence over
those multipliers, with a real-time-feasibility cost per sample, while
the chain resolves the same constraint conflict one-shot at design
time. Both encode the same physics, namely which inequality yields when
two active constraints conflict, in dual representations. The chain
case keeps the resolution rule visible and traceable in the
orchestrator's source code; the distributed-MPC case absorbs the
resolution rule into the multipliers and exposes it only through the
converged solution. Those two representations are dual. The implication
for an architect adding an agent is therefore well defined: the new
priority slot is a discrete design choice, the RGA and Niederlinski
analyses inherited from the underlying plant tell the architect whether
the candidate pairing is admissible at all, and the literature on
priority shifts under active-constraint changes~\cite{reyes2019,reyes2020}
provides the formal vocabulary for ranking the new agent against the
existing ones.

The third limitation is that the LLM is not capable of rescuing a
poorly designed chain. The operator agents in Architecture~C emit a
3-class label (\texttt{IDLE}, \texttt{ACTIVE}, \texttt{SATURATED})
together with a bounded rationale string, and a fixed schema rejects
any output that does not conform to that
structure~\cite{anthropic-docs}. The information capacity of the
discrete-mode channel is therefore at most
$\log_2 3 \approx 1.58\,$bit per call. A future continuous-intensity
output (the Lever~C extension flagged in §3.6) would widen the channel
by $\log_2 K_\text{intensity}\,$bits per call, where $K_\text{intensity}$
is the effectively-resolvable number of intensity quantiles after the
slew clamp, which puts $K_\text{intensity}$ in the 4--8 range and the
total channel ceiling in the 3.5--4.5 bit/call range. The architectural
decision space (CV--MV pairing, RGA selection, priority order across the
selector agents) lives in a much higher-dimensional discrete space: a
plant-wide problem with $n$ controlled variables has $n!$ candidate
pairings, and the cardinality of the priority-rank set on $m$ agents
grows as $m!$ in turn. It is possible to see that encoding a re-pairing
decision in the operator agent's output schema is structurally
impossible at any of those bit budgets: the channel does not have the
bits, and widening the channel further would also widen the
hand-validated set of safe actions, defeating the safety argument that
motivates the schema in the first place.

The same argument has a token-budget restatement. The Qwen~2.5 7B
backbone used by the operator agents has a $32{,}768$-token context
window; the snapshot the orchestrator hands to each operator agent each
sample includes the recent measurement history, the active priority
structure, and the rationale of the previous decision. Even if the
schema were widened to admit a structural recommendation, the context
budget is too small to enumerate the $n!$ candidate pairings that a
plant-wide RGA analysis would consider. The LLM is therefore an
inherently local-decision agent, and a global-architecture agent would
have to be of a different kind: either a different model class trained
on the architectural search space or a deterministic combinatorial
search wrapped around the chain. The information channel exposed by
the present design is sized for parameter scheduling at the regulatory
tier, not for architecture selection.

Two additional observations sharpen the same point. A plant whose
chain is mis-paired produces snapshots the LLM has not seen in any
plausible training corpus, so its rationale-text in those regimes is
unconstrained extrapolation, not retrieval. Recent results on
the theoretical limits of embedding-based retrieval show that an LLM is
not capable of retrieving facts that are absent from its training
distribution~\cite{weller2025rag}, and the affordance argument from the
SayCan line of work makes the matching point on the action side: an
LLM-driven controller is bounded by the environmental affordances
exposed to it, not by the claims its rationale-text is willing to
make~\cite{ahn2022saycan}. A consequence of that bound is exploited
by construction in Architecture~C: the rationale string is logged to
JSONL for the auditor and is read by humans only, while the
deterministic chain consumes only the schema-validated \texttt{mode}
field and the deterministically-computed $u_\text{request}$ scalar. A
hallucinated rationale therefore never reaches the plant, and the
0-fallthrough record observed over the 4-day scenario is the empirical
expression of that property. The agentic-framework deployment of an LLM
at the supervisory tier of an industrial
process~\cite{industrial_llm_agentic2024} relies on the same
combination of schema-bounded outputs and human-validated affordances,
and our Architecture~C reproduces that pattern.

The fourth limitation, taken together with the previous three, is that
the design grammar of an LLM-supervised selector chain is
process-control-theoretic, and not agent-theoretic. The primitives that
decide whether a given plant admits an LLM-supervised chain solution
are inherited from the classical literature on selector control and
plant-wide design, and they precede any choice of agent boundary,
message contract, or orchestrator policy. They are the following.

\begin{itemize}
\item \emph{CV--MV pairing.} Which controlled variable is to be
regulated by which manipulated variable, decided through the RGA
$\lambda_{ij}$ of Bristol~\cite{skogestad2005}, with the recommendation
$\lambda_{ij}$ near unity and no negative diagonal entries.

\item \emph{Interaction analysis.} Whether the chosen pairing is
structurally stable under integral action, tested through the
Niederlinski Index~\cite{niederlinski1971}. A negative NI rules out the
candidate pairing at design time, before any chain is built around
it.

\item \emph{Constraint priority.} Which inequality each agent defends
and which is allowed to yield, decided from the active-set analysis of
the plant under the disturbance scenarios of interest. The formal
treatment of priority shifts as the active set changes is the subject
of recent work on selector control and split-range
generalisations~\cite{reyes2019,reyes2020,forsman2025splitparallel}.

\item \emph{Plant-wide procedure.} A degrees-of-freedom analysis,
followed by an economic-objective statement, CV selection, MV
selection, throughput-manipulator selection, and a stepwise
control-structure design, is the canonical sequence of decisions for a
plant-wide problem~\cite{skogestad2000plantwide,skogestad2026arc}. The
selector chain in this paper is the regulatory layer of that
sequence, applied to the cow-barn case study; the same sequence
applies to any plant where an LLM-supervised chain is a candidate.

\item \emph{Selector primitives.} The MIN, MAX, override, split-range,
and mid-ranging primitives of the classical selector-control
literature~\cite{maarleveld1970,shinskey1981,forsman2025splitparallel,reyes2020}
are the building blocks from which the orchestrator's resolution logic
is assembled. Once the pairing and priorities are fixed, the choice
among those primitives is a transcription, and not a design step.

\item \emph{Selector synthesis.} Whether each layer of the chain uses
a MIN or a MAX is determined by Skogestad's Selector Rules 1 and 2: a
constraint satisfied with a large MV is defended by a max-selector, a
constraint satisfied with a small MV by a min-selector, and conflicting
cases require max and min selectors in series with $u_0$ entering the
first selector~\cite{skogestad2023arc,skogestad2026arc}. Those two
rules are the chain-synthesis step that turns the priority ranking
into a wired network of MIN and MAX nodes.
\end{itemize}

Those primitives are decided before the agent boundary is drawn and
they are not capable of being rescued by any subsequent choice of
orchestrator policy or LLM advisor. The agent boundary, the message
contract, and the workspace discipline are implementation conveniences
that follow from the priority structure; they carry the architecture
but they are not the engine of correctness. Two orthogonal routes are
available to an architect who wishes to inject control-theoretic
knowledge into an LLM-supervised chain: the first is through the
LLM's input schema, where the state-summary prompt encodes the active
priority order, the recent measurement history, and the regime context;
the second is through the deterministic chain that wraps the LLM,
where the priority structure, the selector primitives, and the override
network encode the same knowledge in a form the LLM never sees. The
two routes carry complementary slices of the same control-theoretic
content, and a sound design uses both. The route through the chain
carries the architecture; the route through the schema carries the
regime-aware parameter scheduling. Practitioners who treat the agent
system as primary and the chain as a wrapper recover, in our
preliminary experience, agent boundaries that are difficult to align
with the underlying constraint geometry.

Those four limitations are not unique to our implementation; they are
structural features of any chain-based multi-agent process-control
design that places a language model at the supervisory tier. It is also
useful to ground each of them back to the cow-barn deployment, because
the numbers the case study produced are the empirical anchor for the
arguments above. The Amdahl scaling at $N=6$ ($S(6)\approx 5$) is
comfortably below the per-sample budget set by the 5-minute polling
cadence: a single Qwen 2.5 7B call is in the 100--500\,ms range on the
test workstation, so the per-sample wall-clock cost of the six operator
agents sits inside the 5-minute budget with two orders of magnitude of
margin even before batching. The RGA computed above on the 2$\times$2
plant supports the C-with-fan and T-with-heater pairing the chain
actually implements ($\lambda_{11}=\lambda_{22}=1.00$ at the nominal),
so the structural-pairing primitive is exercised on this plant rather
than asserted. The information-theoretic ceiling shows up empirically:
across the 4-day mixed-season scenario the six operator agents made
6\,912 schema-validated calls (six agents, 1\,152 calls each), 0 of
them produced a fallthrough, and the priority order between
\texttt{tc\_0} and \texttt{cc\_1000} was never altered by the LLM, as
the information channel does not have the bits to encode that
re-pairing decision. The design-grammar primitives, finally, are the
language in which the cow-barn chain was assembled before any LLM
was introduced; what the LLM contributed in this paper was confined to
the parameter-scheduling tier. The architectural know-how of selecting
agents, ranking their priorities, and partitioning the constraint set
across manipulated variables remains a human task; the language model
contributes only at the parameter-scheduling tier, and the four
limitations stated above are the structural reasons why that remains
the right division of labour.

\section{Conclusions}
\label{sec:conclusions}

This work proposed a multi-agent reformulation of the regulatory control
layer that is derived from Advanced Regulatory Control theory. The starting
point is the observation, supported by the recent literature, that large
language models are useful tools for general-purpose tasks yet present poor
performance on specific domain tasks, partly because supplying narrow
context to a general-purpose model is difficult and the task being asked of
the model is rarely bounded by construction. Process control theory, and
ARC in particular, offers a discipline for decomposing a regulatory problem
into elements of contained scope, each defending one inequality on one
controlled variable, with conflicts resolved by structural priority through
MIN/MAX selector networks. The hypothesis explored here is that this same
discipline informs the construction of safe multi-agent LLM systems for
process control.

The hypothesis is supported by the case studies. ARC theory is capable of
providing a principled recipe for the construction of a multi-agent system
at the regulatory tier: the agent roster is derived from the plant-wide
procedure under two structural equivalences, each PI in the chain
becoming one operator agent and the chain's interaction logic becoming
one orchestrator agent. Those points are what supports the structural
soundness of the architecture, since every constraint conflict is resolved
deterministically by the orchestrator following the priority order
prescribed by control theory, regardless of what the LLM operator agents
emit.

A consequence of the agent reformulation is that the PI tuning step of
classical ARC is replaced by a prompt engineering and context curation
step. The LLM operator agent does not require numerical gain tuning, since
it produces a regime classification that the agent code maps to a proposal
through the agent's own identity. What the operator agent does require is
a curated context, namely its role in the chain, the selector kind, the
binding direction, and the rolling history of its own past decisions.
This trade-off is the design choice that the practitioner adopting the
present architecture inherits. The PI tuning literature is mature, while
the prompt engineering for regulatory-tier agents is at an earlier stage,
and the storytelling of the prompt design in this work is intended to
inform that practice.

The Claude-based LLM orchestrator was also tested. It is possible to wire
a language model into the supervisory tier of the chain with bounded
outputs, schema validation, and a deterministic fallback, and the
orchestrator-tier LLM is capable of varying the chain's tuning parameters
within regime classifications, not between them. The implication is
that the language model's contribution at the supervisory tier is in
continuous-parameter retuning, while the discrete priority order remains
under the deterministic chain. The two roles, operator at the regulatory
tier and orchestrator at the supervisory tier, occupy different tiers of
the plant-wide hierarchy and informs how an LLM may be deployed at each.

Several limitations remain. The operator agents in this work are local
instruction-tuned models running offline, motivated by data access
restrictions, security considerations, and the absence of cloud
connectivity at many process sites. Local models inform a deployment style
that is feasible in industrial settings but constrains model capacity, and
the prompt engineering effort observed here is one of the costs of that
choice. A wider validation, including Monte Carlo replication, regime
sweeps beyond the cold regime and mixed-season conditions, and field
deployment, is needed before a preference among architectures can be
asserted; those points are explicitly deferred to future work.


\end{document}